# Observation of negative surface and interface energies of quantum dots

Jason J. Calvin[1,2], Amanda S. Brewer[1,2], Michelle F. Crook[1,2], Tierni M. Kaufman[1], and A. Paul Alivisatos[1,2,3,4]*

[1]Department of Chemistry, University of California, Berkeley; Berkeley, California 94720, United States

[2]Material Sciences Division, Lawrence Berkeley National Laboratory; Berkeley, California 94720, United States

[3]Department of Materials Science and Engineering, University of California, Berkeley; Berkeley, California 94720, United States

[4]Kavli Energy NanoScience Institute; Berkeley, California 94720, United States

*A. Paul Alivisatos

Email: paul.alivisatos@berkeley.edu

**Abstract**

Surface energy is a fundamental property of materials and is particularly important in describing nanomaterials where atoms or molecules at the surface constitute a large fraction of the material. Traditionally, surface energy is considered to be a positive quantity, where atoms or molecules at the surface are less thermodynamically stable than their counterparts in the interior of the material because they have fewer bonds or interactions at the surface. Using calorimetric methods, we show that the surface energy is negative in some prototypical colloidal semiconductor nanocrystals, or quantum dots with organic ligand coatings. This implies that the surface atoms are more thermodynamically stable than those on the interior due to the strong bonds between these atoms and surfactant molecules, or ligands, that coat their surface. In addition, we extend this work to core/shell indium phosphide/zinc sulfide nanocrystals and show that the interfacial energy between these materials is highly thermodynamically favorable in spite of their large lattice mismatch. This work challenges many of the assumptions that have guided thinking about colloidal nanomaterial thermodynamics, investigates the fundamental stability of many technologically relevant colloidal nanomaterials, and paves the way for future experimental and theoretical work on nanocrystal thermodynamics.

**Significance Statement**
As materials approach the nanoscale, a significant portion of their atoms reside on their surface. Quantum dots are semiconductor nanocrystals that have found applications ranging from displays to solar cells, and it has been broadly assumed that their surface atoms are less stable than atoms in the interior of the crystal, which is described as having a positive surface energy. However, in this work we use calorimetry to show that the surface energy of several different quantum dots is negative, meaning that the surface atoms are more stable than the interior atoms. We attribute this negative surface energy to tightly bound surface molecules. This result changes our fundamental understanding of the thermodynamic stability of these technologically relevant materials.

**Main Text**

**Introduction**

As materials approach the nanoscale, surface energy plays an increasingly important role in their overall thermodynamic stability (1-4). Almost invariably, surface energy, denoted as $\gamma$, is considered to be a positive quantity. Atoms or molecules at the surface of a material are widely considered to be in a less thermodynamically stable state than those in the interior, and this is attributed to broken bonds or fewer favorable interactions at the surface (5, 6). In general, this would drive materials to adopt low surface area configurations, minimizing the overall energy of the system (4, 7). At the nanoscale, this size-dependent effect of surface energy on material stability can be so impactful that it has been observed to invert the phase stability in titania. While rutile is the most thermodynamically stable phase in bulk, nanocrystalline anatase is more thermodynamically stable than nanocrystalline rutile because it has a lower surface energy (2).

Nanocrystals often have organic species adsorbed to their surface as a deliberate design choice with many advantages, and these species *will influence their surface energies* (2-4, 8-14). These species are often referred to as ligands, and the structure of molecules adsorbed at nanocrystal surfaces, their binding motifs, and their impact on nanocrystal properties cover a wide range of research in the nanocrystal field (10, 15-17). These organic ligands impact the growth (15, 18-20), colloidal stability (16, 21), optical and electronic properties (12, 17, 22), self-assembly into superlattices (16, 23), and nanomaterial applications (24, 25). Since organic ligands serve as stabilizing agents for atoms at the surface, they should substantially lower the surface energy of a material, compared to a bare nanocrystal in vacuum (3, 4, 8, 11, 26, 27).

The surface energy will reflect the detailed atomic structure, including faceting, edges, and corners and their various reconstructions. However, simple models using a single surface energy to represent a composite of these factors are often sufficient to explain the overall thermodynamic trends of materials in the two to twenty nanometer length scale (1, 2, 7). Surface energy has been used to justify why nanocrystals adopt certain shapes, often using Wulff constructions (7, 28), and their growth mechanisms (7, 11, 18-20, 29). As a result, substantial effort has been devoted to calculating the surface energies of crystal facets for a variety of materials, with and without ligands (11, 18, 19, 27). While previous calorimetric measurements have investigated nanocrystals with bare or water coated surfaces (1-3, 30, 31), to our knowledge, there are no calorimetric measurements of these fundamental quantities that determine the relative stabilities and shapes of nanocrystals coated with organic ligands or encapsulated in an inorganic shell.

For colloidal semiconductor nanocrystals, or quantum dots, these ligands often take the form of molecules that consist of a head group, such as a carboxylate, that strongly coordinates to the surface of the quantum dot, as well as a tail group like an aliphatic chain (8, 10, 15, 17, 22, 27). These components play important roles in dictating the properties of quantum dots, where the head group can "passivate" electronic defects on their surface by pulling these levels out of the band gap (12, 17, 22). The tail group prevents quantum dot fusion during synthesis and allows for colloidal stability in organic solvents (16, 21). A wide variety of ligands can be used with these colloidal quantum dots and can be tailored to change a quantum dot's physical and chemical properties and to introduce them into a wide range of environments. The organic-inorganic interface of colloidal quantum dots is a large part of what makes them interesting objects of study and useful in technologies. Hence measuring the energy associated with these interfaces is foundational to the field.

**Results and Discussion**

To study the thermodynamics of the surfaces of colloidal quantum dots, we measured the surface energy of quantum dots capped with oleate ($C_{18}H_{33}O_2$) using dissolution calorimetry. This technique has been well established to measure the surface energies of nanocrystals (2, 3, 26), but, to the best of our knowledge, has never applied to colloidal quantum dots. Changing the size of the nanocrystal modulates the proportion of atoms on the surface of the nanocrystal versus those in the interior. By dissolving these materials into atomic species and monitoring the heat flow during the reaction, the enthalpy of the reaction can be calculated. Fig. 1A illustrates the experimental method for this technique. Indium phosphide, either as a bulk material or as agglomerated quantum dots deposited on a Teflon platform, is dissolved in hydrochloric acid in a sample chamber, and the differential heat flow between the sample chamber and a reference chamber filled with hydrochloric acid is recorded. The sample and reference chambers are maintained at an isothermal temperature of 25 °C. The heat flow during the reaction is depicted in Fig. 1B and 1C for the bulk indium phosphide and quantum dots, respectively. The heat flow is integrated over the reaction time to determine the enthalpy of the reaction, and the enthalpy of the dissolution of the quantum dots is corrected by subtracting the enthalpy of dropping the bare Teflon platform and the protonation enthalpy of the oleate ligands. If the enthalpy of dissolution of the material in terms of kJ/mol is plotted versus the molar surface area ($m^2$/mol), or surface area in $m^2$ per mole of material, the slope of that line represents the excess enthalpy of the reaction as a function of surface area ($J/m^2$). The negative of that slope represents the excess enthalpy of the reverse reaction (3), or the formation of nanocrystals from atomic species as a function of surface area. This excess enthalpy is defined as the surface enthalpy of a material. Rigorously, this calorimetric method measures enthalpy and there is an entropic contribution to the Gibbs free energy of formation of these surfaces, or a surface Gibbs free energy (31). However, this entropic contribution has been measured by low temperature heat capacity measurements on magnesium and cobalt oxide nanoparticles and has been found to be quite small compared with measured surface energies, on the order of 0.3 mJ/K·$m^2$ at 298.15 K (TS = 0.08 $J/m^2$) which is much smaller than measured surface enthalpies (30, 31). For this reason, the measured surface enthalpy is an excellent proxy for surface energy (1-3). Note that these results are not facet dependent surface energies for the quantum dots as these quasi-spherical quantum dot surfaces are not well defined, but rather an average surface energy for the quantum dots.

Fig. 2 depicts the enthalpy measured by dissolution calorimetry experiments plotted against the molar surface area for a size series of indium phosphide quantum dots. The molar surface area was determined using scanning transmission electron microscopy (STEM) imaging (extensive sample characterization can be found in *SI Appendix*, Fig. S1 as well as Tables S1–3). A critical aspect of these measurements is ensuring surfaces that are chemically well-defined. As indium phosphide is particularly susceptible to oxidation, even during the synthetic process, we employed a recently developed hydrogen atmosphere synthesis (10, 32). No surface oxidation was detected by solid state NMR (see *SI Appendix*, Fig. S1F and S1G). Unlike past measurements of nanocrystal surface energies (3), the slope of this plot is *positive*, indicating that the surface energy of these indium phosphide quantum dots is *negative*, or that the ligand-bound atoms on the surface of the quantum dots are more stable than the atoms in the interior. This result is highly unusual, as surface energies, while never actually measured for these colloidal quantum dots, have almost universally been described as possessing positive values (6, 8, 11, 27).

To investigate the generality of this phenomenon, size series of zinc sulfide, cadmium sulfide, and lead sulfide quantum dots capped with oleate were synthesized and interrogated via dissolution calorimetry, and the results are depicted in Fig. 3 (extensive sample characterization can be found in *SI Appendix*, Fig. S1–4 as well as Tables S1–3). For each of these materials, the surface energy was negative, although less negative than that of indium phosphide. This indicates that a negative surface energy may be a common property of colloidal semiconductor quantum dots capped with carboxylate ligands and is not unique to indium phosphide quantum dots.

Although it is well accepted that surface species, such as oleate ligands, lower the surface energy of a material (3, 8, 11, 14, 27), there has been substantial controversy regarding whether the surface energy of a material can be lowered enough to become negative. A negative surface energy implies that smaller nanocrystals are more thermodynamically stable than larger ones, so that nanocrystals ought to disintegrate in the long-time thermodynamic limit, and this implication has caused concern (6). For example, a DFT study claimed that negative surface energies could be possible on alumina with strongly adsorbed water (26), and additional theoretical studies have described this phenomena in gold nanocrystals capped with thiols and platinum nanocrystals with adsorbed carbon monoxide as well (33-35). The publication of the study on alumina reportedly "caused confusion in the scientific community" and prompted a follow up correspondence that called the use of the phrase negative surface energy "provocative" but that it is "possible to predict positive and negative surface energies for any number of multicomponent systems…" (36). As identified by that correspondence, one cause of this controversy stems from different interpretations of the definition of surface energy, depending on the reference to the vacuum. If the assumption is that atoms at a surface have fewer bonds than atoms in the interior, then it follows that the surface energy will be positive (Fig. 4A). However, the presence of ligands means that surface atoms do not necessarily have dangling bonds, and if the surface ligands bind more tightly to the surface atoms than their interior bonds, the material will therefore have a negative surface energy (Fig. 4B). It may be more appropriate to refer to this quantity not as a surface energy, but perhaps an interface energy, where the thermodynamic stability of the interface between the nanocrystal and the adsorbed species is being studied. The interfacial energy would then be the sum of the surface energy of a bare nanocrystal surface plus the energy of adsorption of the surface species (7, 19). Although a negative interfacial energy naively implies that nanocrystals are not in their most thermodynamically stable state, metastable systems certainly exist and persist for meaningful time periods, even centuries or geologic timescales. Indeed, the existence of a positive surface energy naively suggests that larger nanocrystals are more stable, and that any size of nanocrystal is therefore also in a metastable state with respect to the bulk, yet it is widely accepted that the kinetic barriers to fusion of particles can be engineered to persist on nearly any timescale. However, it has been suggested that a size dependent positive interfacial energy may result in nanoscale systems that are thermodynamically stable (37), and this could also be true for negative interfacial energy systems, as was suggested studies that observed the growth of zinc sulfide nanosheets in strongly basic conditions, though no direct measurements of the interfacial energy was done (38-40).

While the vast majority of literature assumes that "surface energies" must be positive, there are a few reports of negative surface stresses, negative surface tension, and negative surface energy, although these have exclusively been reported to the field of experimental transmission electron diffraction and X-ray diffraction (4, 9, 13, 41, 42). With these studies, surface stress can be defined as

$$f_{ij} = \gamma \delta_{ij} + \frac{\partial \gamma}{\partial \varepsilon_{ij}}$$

such that $f_{ij}$ is the surface stress, $\gamma$ is the surface energy, $\delta_{ij}$ is the Kronecker delta, and $\varepsilon_{ij}$ is the surface elastic strain tensor, where i,j = 1,2 (5, 43). This is a generalized version of the definition originally put forth by Shuttelworth (44, 45), but unfortunately, there is significant controversy surrounding the exactness of that definition (46-48). Regardless, surface energy and surface stress are not identical quantities. The derivative term can be positive or negative, and hence the surface stress can be positive or negative while avoiding the assertion that the surface energy is negative (5, 43, 45). In X-ray diffraction studies, only the surface stress can be measured, although the terms surface stress, surface tension, and surface energy are often used interchangeably (4). Arguably, the term surface tension should be avoided for solids (43), but it is pervasive in the literature (9, 13, 30, 44, 48). On the other hand, calorimetric measurements can directly measure the surface energy, and are therefore more appropriate for surface thermodynamics studies. While direct calorimetric measurements on bare or water capped nanocrystals have yielded positive surface energies (1-3, 30, 31), indirect measurements have indicated the possibility of a negative surface energy for nanocrystals capped with organic ligands. Multiple studies have attributed this surface stress to the binding of ligands, and it is interesting that the measured surface tension by Bertolotti *et al.* for lead sulfide nanocrystals capped with oleate was -0.618 J/m$^2$ and -0.537 J/m$^2$ in different publications (9, 13), and the estimated "surface energy" by Zhao *et al.* for cesium lead iodide nanocrystals capped with oleic acid and oleylamine was between -0.48 and -0.82 J/m$^2$ (4). Discrepancies between measurements herein and those by Bertolotii *et al.* may be attributed to the difference between surface energy and surface stress, but could also arise from differences in synthetic methods or ligand coverages.

From our initial data, we hypothesized that negative surface energies originate from the tightly bound ligands on the quantum dot surfaces. It is well known that in carboxylate terminated quantum dots the carboxylate coordinates to metal cations on the surface, and if the oxygens from the carboxylates were binding more tightly to the metal cations on the surface than the underlying inorganic crystal lattice, this would result in a negative surface energy (Fig. 4B). Further, as the group III indium likely has a nearly trivalent oxidation state, the carboxylate likely coordinates more strongly to indium than to group II and IV "divalent" cadmium, zinc, or lead, potentially explaining why the surface energy of indium phosphide is more negative than that of the other quantum dots. A logical extension of our hypothesis is that if the ligands coordinating the surface of the nanocrystal had similar bond strength to the underlying inorganic crystal lattice, the measured surface energy would be approximately zero (Fig. 4C). While we hypothesize that this would be true for a variety of metal oxide nanocrystals capped with carboxylate ligands, to validate this prediction, we sought a metal oxide nanocrystal capped with the same ligands as the quantum dots and that could be measured under identical experimental conditions as used for the quantum dots. To test this, we therefore synthesized cadmium oxide nanocrystals capped with oleates and measured their dissolution enthalpy (extensive sample characterization can be found in *SI Appendix*, Fig. S5 as well as Tables S1–3). As expected, the surface energy of these nanocrystals was close to zero (Fig. 4D), indicating that the atoms on the surface and interior had similar bond enthalpies, which is reasonable considering oxygen was coordinating to cadmium on the exterior and interior.

Having established a framework to measure the surface energies of colloidal quantum dots, we were able to extend this work to also measure the energy of the interface of a core/shell nanocrystal heterostructure. While organic ligands can be used to partially passivate defects on the surface of colloidal quantum dots to improve their luminescence, the growth of a wider band gap material to coat the surface of the core quantum dot, or a shell, has resulted in nanocrystals with near unity quantum yields that are better suited for optoelectronic applications (49-51). For this measurement, indium phosphide/zinc sulfide core shell structures were synthesized with a variety of core sizes, resulting in different molar interfacial areas (extensive sample characterization can be found in *SI Appendix*, Fig. S1 and S6 as well as Tables S1–3). The samples were dissolved in the same manner, and the resulting enthalpies were corrected for the zinc sulfide content and the surface energy of the zinc sulfide shell, which were calculated from the surface energy measurements of the zinc sulfide quantum dots, as well as the protonation enthalpy of the oleate ligands on the zinc sulfide shell. After accounting for these enthalpy contributions, the remaining enthalpy reflects the enthalpy of dissolution of the indium phosphide core, and the interfacial energy between the indium phosphide core and the zinc sulfide shell can be calculated using these values. Fig. 5 depicts the results of these measurements, which show a negative interface energy between indium phosphide and zinc sulfide. These measurements indicate that the formation of the zinc sulfide shell on the indium phosphide quantum dot is highly favorable, stabilizing the surface indium atoms much more significantly than the carboxylate ligands. As with the surface energy measurements, this negative interface energy result is unexpected and surprised us. Beyond the convenience of using materials that had already been studied as the core and shell material and their relevance as more environmentally friendly quantum dot materials (49, 50), indium phosphide and zinc sulfide were selected for this study precisely due to their large lattice mismatch. We anticipated that this would result in a positive interfacial energy due to lattice strain (52-54); however, the bond enthalpy between these materials must be much more favorable than the lattice strain to result in such a negative interfacial energy. In retrospect, the existence of core/shell heterostructures indicates that the formation of the interface must be favorable, else the separate formation of nanocrystals consisting only of the shelling material or Janus type nanostructures, rather than a coating of the core quantum dot, would be more likely. While a negative interfacial energy may predict the formation of certain heterostructures that maximize interfacial area at thermodynamic equilibrium, kinetics, sterics, and activation energies are also important in dictating the structures formed under non-equilibrium conditions. These measurements also have interesting parallels with anion exchange in cadmium chalcogenide nanocrystals, where forming an alloy or an interface depends on the identity of the exchanging cation (55). Using zinc or mercury ions forms an alloy, while lead forms interfacial structures (56-58). The difference between forming alloys or interfaces is likely affected by lattice constants, ion size, crystal structure, and entropy (55), but the interfacial energy between these materials almost certainly impact what structure is formed. These interfacial energy measurements could be used to further explore the formation of these nanoheterostructures. Strong chemical bonds between different materials with different lattices also impacts the field of catalysis, where strong metal–support interactions can decrease catalytic efficiency (59, 60), and this technique provides an opportunity to directly probe these interactions.

From cross polarized $^1$H-$^{31}$P magic-angle spinning solid state NMR measurements, it is evident that the indium phosphide quantum dots are completely coated in a zinc sulfide shell (see *SI Appendix*, Fig. S6F). Beyond the initial coating, however, the strain energy in the shell would likely result in a non-uniform shell, which has been observed in these systems as well (49, 53, 54, 61). In many ways, this is similar to atomic layer deposition of materials on a substrate, where often two-dimensional growth is first observed before the growth of islands to relieve lattice strain (62-64). For this particular heterostructure, it is plausible that sulfur anions coordinate to the indium rich surface as the shell is grown, but the exact chemical nature of this interface is still ambiguous (65-67). Particularly from a charge balance perspective, simple growth of a zinc blende zinc sulfide shell would not passivate a zinc blende indium phosphide surface. One possibility is that extra sulfur atoms at the interface maintain charge balance while contributing to the large negative interfacial energy observed. Regardless of the exact nature of the interface, the results unambiguously indicate that the formation of these core/shell structures is more thermodynamically stable than separated nanocrystals or bulk materials.

From these measurements, we have experimentally demonstrated that surface energies of colloidal nanoparticles can be negative, and the assumption that surface energies should be positive values may be incorrect when accounting for molecules binding to the surface. The implications of these results may require the re-evaluation of previous work assuming positive surface energies, particularly for growth models of nanocrystals. The most common model for nanocrystal nucleation, the LaMer model (29), assumes that the surface energy is positive. A negative surface energy implies that smaller nanocrystals are more favorable than larger nanocrystals, and in the thermodynamic limit, bulk inorganic crystals and ligands would favor the formation of nanocrystals. It is important to note that the surface energies of these quantum dots were measured at 298.15 K, but nanocrystal nucleation often occurs at higher temperatures, and it is likely that the surface energy could be temperature dependent where organic ligands may not bind as tightly or at the density of the ligands at room temperature. Additionally, entropy contributions at higher temperatures will be more significant, particularly with ligands that can be free in solution or bound to the nanocrystal. We also know that the linear trend of formation enthalpy with size cannot continue indefinitely due to their finite nature, as nanocrystals become molecular clusters of atoms as size decreases beyond that of the particles measured herein (68, 69), and we expect that the thermodynamics of cluster formation to deviate from this behavior. Furthermore, the thermodynamics of nanocrystal formation are also likely driven by the thermodynamic chemical potentials of starting materials, such as reactive phosphorus and sulfur precursors, and this should be a rich area of future study.

The observation of negative surface energies and a negative interface energy in indium phosphide/zinc sulfide core/shell nanocrystals has wide implications, from practical considerations to interesting new directions in experimental and theoretical materials science. First and foremost, these negative values indicate that these nanocrystals are more thermodynamically stable than their bulk counterparts. This is supportive of findings on unstable materials, such as lead halide perovskites (70-72), that appear to be more stable as nanoparticles than in bulk (4). Second, the vast body of literature that calculates surface energies of nanoparticles or uses surface energy as a justification for observed phenomena where it is implicitly assumed that surface energy must be positive must be re-evaluated. Third, the elimination of what was considered to be a fundamental assumption opens the door to new theories regarding the stability of nanomaterials and possible materials that can be formed. We expect that this technique of measuring the surface energy of nanoparticles will be widely adopted to further investigate the field of experimental thermodynamics of nanocrystals, particularly regarding colloidal nanocrystals.

## Materials and Methods

### Materials

Cadmium oxide (99.99%), oleic acid (technical grade, 90%), and n-hexadecane (99%) were purchased from Sigma Aldrich and used as received. Indium acetate (99.99%), diethyl zinc (≥52 wt % Zn basis), cadmium acetate (99.995%), cadmium acetylacetonate (≥99.9%), lead oxide (>99.9%), oleic acid (≥99%), anhydrous hexanes (>99%), anhydrous methyl acetate (99.5%), and toluene-$d_8$ (99.6%) were purchased from Sigma-Aldrich and stored in an argon glovebox. Anhydrous tetrahydrofuran (≥99.9%, Sigma-Aldrich) was distilled from a solvent still immediately prior to being stored in an argon glovebox. Indium (99.9999%), zinc (99.9999%), cadmium (99.999%), indium phosphide (99.9999%), zinc sulfide (99.995%), and cadmium sulfide (99.999%) were purchased from Alfa Aesar and used as received. Lead sulfide (99.995%) and hydrochloric acid (37%, ACS Reagent) were purchased from Fisher Scientific and used as received. Argon (99.999%), and hydrogen (5% in argon) were purchased from Airgas and used as received. Zinc acetate (99.98%, Alfa Aesar), tris(trimethylsilyl)phosphine (98%, Strem Chemicals), bis(trimethylsilyl)sulfide (98%, Beantown Chemical), bis(trimethylsilyl)amine (98%, Acros Organics), and anhydrous acetone (99.5%, Fisher Scientific) were stored in an argon glovebox and used as received.

### Synthesis of indium phosphide quantum dots

For the synthesis, 0.350 grams (1.20 mmol) of indium acetate with 3.60 mmol of 90% oleic acid was combined with 10 mL of *n*-hexadecane in an oven-dried, 50 mL three neck round-bottom flask with a Teflon coated stir bar. After an air condenser column, septa, and thermocouple adapter were attached to the flask and all glass joints greased with Apiezon H grease, the flask was evacuated via Schlenk line. The temperature of the flask was raised slowly to 70 °C, after which the vacuum line was closed. The solution was slowly heated to 80 °C, and the line to the vacuum was then opened slightly. The solution was then slowly heated to 120 °C, and all solids disappeared, resulting in a clear solution. Maximum temperature reached by the solution tended to be 130 °C. The temperature was lowered to 80 °C for *n*-hexadecane after which the vacuum line was fully opened, reaching a pressure below 60 mTorr. The system was flushed with 5% hydrogen in argon gas and placed back under vacuum, and this process was repeated three times in intervals of ten minutes. In total, the degassing process lasted one hour. The system was filled with 5% hydrogen in argon and the temperature was raised to 110 °C. In an argon glovebox, 1.540 g of *n*-hexadecane that had previously been heated and degassed for an hour at 90 °C, respectively, at 60 mTorr was combined with 0.152 g of tris(trimethylsilyl)phosphine (0.607 mmol) and loaded into a syringe. This solution was then swiftly injected into the flask, and temperature was maintained for two minutes, resulting in a yellow solution. The temperature was then raised to 190 °C at a rate of 20 °C/minute and maintained at this temperature between 2 and 45 minutes. The solution turned a deep red to completely black depending on the length of the growth period, and then the flask was then rapidly cooled under a stream of nitrogen and hexanes. After cooling the solution was cannulated to be purified in an argon glovebox.

**Synthesis of zinc sulfide quantum dots**

In an oven-dried, three-neck, 50 mL round-bottom flask, 0.220 g of zinc acetate (1.2 mmol) was added to 10 mL of *n*-hexadecane and 0.678 g of technical grade oleic acid (2.4 mmol) with a Teflon-coated stir bar. A Schlenk line was attached to a condenser column on the middle neck, and the other necks were sealed with a thermocouple adapter and a septa cap. Joints were sealed with Apiezon H grease and the flask was evacuated via Schlenk line. The temperature of the flask was raised slowly to 70 °C, after which the vacuum line was closed. The solution was slowly heated to 80 °C, and the line to the vacuum was then opened slightly. The solution was then slowly heated to 120 °C until all solids disappeared, resulting in a clear solution. The temperature was lowered to 80 °C after which the vacuum line was fully opened, reaching a pressure below 60 mTorr. The system was flushed with 5% hydrogen in argon gas and placed back under vacuum, and this process was repeated three times in intervals of ten minutes. In total, the degassing process lasted one hour. Then, the solution was placed back under 5% hydrogen in argon, and the temperature was raised to 250 °C at a rate of 20 °C per minute. In an argon glovebox, 1.540 g of *n*-hexadecane that had previously been heated and degassed for an hour at 90 °C at 60 mTorr was combined with 0.108 g of bis(trimethylsilyl)sulfide (0.6 mmol) and loaded into a syringe. This solution was then swiftly injected into the flask, and temperature was maintained for five minutes. The flask was then rapidly cooled under a stream of nitrogen and hexanes. After cooling the solution was cannulated to be purified in an argon glovebox.

**Synthesis of cadmium sulfide quantum dots**

In an oven-dried, three-neck, 50 mL round-bottom flask, 0.278 g of cadmium acetate (1.2 mmol) was added to 10 mL of *n*-hexadecane and 0.678 g of technical grade oleic acid (2.4 mmol) with a Teflon-coated stir bar. A Schlenk line was attached to a condenser column on the middle neck, and the other necks were sealed with a thermocouple adapter and a septa cap. Joints were sealed with Apiezon H grease and the flask was evacuated via Schlenk line. The temperature of the flask was raised slowly to 70 °C, after which the vacuum line was closed. The solution was slowly heated to 80 °C, and the line to the vacuum was then opened slightly. The solution was then slowly heated to 120 °C, the vacuum was closed, and then the solution was heated to 150 °C until all solids disappeared, resulting in a clear solution. The temperature was lowered to 80 °C after which the vacuum line was fully opened, reaching a pressure below 60 mTorr. The system was flushed with argon gas and placed back under vacuum, and this process was repeated three times in intervals of ten minutes. In total, the degassing process lasted one hour. Then, the solution was placed back under argon, and the temperature was raised to between 150 and 270 °C at a rate of 20 °C per minute. In an argon glovebox, 1.540 g of *n*-hexadecane that had previously been heated and degassed for an hour at 90 °C at 60 mTorr was combined with 0.108 g of bis(trimethylsilyl)sulfide (0.6 mmol) and loaded into a syringe. This solution was then swiftly injected into the flask, and temperature was maintained for five minutes. The flask was then rapidly cooled under a stream of nitrogen and hexanes. After cooling the solution was cannulated to be purified in an argon glovebox.

**Synthesis of lead sulfide quantum dots**

In an oven-dried, three-neck, 50 mL round-bottom flask, 0.268 g of lead oxide (1.2 mmol) was added to 10 mL of *n*-hexadecane and 0.678 g of technical grade oleic acid (2.4 mmol) with a Teflon-coated stir bar. A Schlenk line was attached to a condenser column on the middle neck, and the other necks were sealed with a thermocouple adapter and a septa cap. Joints were sealed with Apiezon H grease and the flask was evacuated via Schlenk line. The temperature of the flask was raised slowly to 70 °C, after which the vacuum line was closed. The solution was slowly heated to 80 °C, and the line to the vacuum was then opened slightly. The solution was then slowly heated to 120 °C, until all solids disappeared, resulting in a clear solution. The temperature was lowered to 80 °C after which the vacuum line was fully opened, reaching a pressure below 60 mTorr. The system was flushed with argon gas and placed back under vacuum, and this process was repeated three times in intervals of ten minutes. In total, the degassing process lasted one hour. Then, the solution was placed back under argon, and the temperature was raised or lowered to between 60 and 180 °C. In an argon glovebox, 1.540 g of *n*-hexadecane that had previously been heated and degassed for an hour at 90 °C at 60 mTorr was combined with 0.108 g of bis(trimethylsilyl)sulfide (0.6 mmol) and loaded into a syringe. This solution was then swiftly injected into the flask, and temperature was maintained for five minutes. The flask was then rapidly cooled under a stream of nitrogen and hexanes. After cooling the solution was cannulated to be purified in an argon glovebox.

**Synthesis of cadmium oxide nanocrystals**

This procedure was adapted from Liu *et al.* (73). In an oven-dried, three-neck, 50 mL round-bottom flask, 0.466 g of cadmium acetylacetonate (1.5 mmol) was added to 10 mL of *n*-hexadecane and 2.4 mL of technical grade oleic acid (7.5 mmol) with a Teflon-coated stir bar. A Schlenk line was attached to a condenser column on the middle neck, and the other necks were sealed with a thermocouple adapter and a septa cap. Joints were sealed with Apiezon H grease and the flask was evacuated via Schlenk line. The temperature of the flask was raised slowly to 80 °C, eventually reaching a pressure below 60 mTorr. In an argon glovebox, 1.540 g of *n*-hexadecane that had previously been heated and degassed for an hour at 90 °C at 60 mTorr was combined with between 0.061 and 0.484 g of bis(trimethylsilyl)amine (0.375 and 3.0 mmol) depending on the synthesis, and this solution was loaded into a syringe. The reaction flask was then flushed with argon, and the bis(trimethylsilyl)amine solution was injected. The temperature was raised to 293 °C, and the solution was vigorously refluxing. The solution was refluxed for between 30 minutes and 3 hours, depending on the synthesis, where higher amounts of bis(trimethylsilyl)amine resulted in quicker syntheses. The reaction was considered finished when the brown solution changed to a dark grey, vaguely greenish color rapidly that coincided with a drop in temperature. The flask was then rapidly cooled under a stream of nitrogen and hexanes. After cooling the solution was cannulated to be purified in an argon glovebox.

**Synthesis of indium phosphide/zinc sulfide core/shell nanocrystals**

In an oven-dried, three-neck, 50 mL round-bottom flask, a Teflon-coated stir bar was added to 10 mL of *n*-hexadecane. A Schlenk line was attached to a condenser column on the middle neck, and the other necks were sealed with a thermocouple adapter and a septa cap. Joints were sealed with Apiezon H grease and the flask was evacuated via Schlenk line. The temperature of the flask was raised to 80 °C, and the flask was evacuated for 30 minutes at a pressure below 60 mTorr. The solution was placed under 5% hydrogen in argon. A calculated amount of indium phosphide quantum dots in hexanes was loaded into a syringe in an argon glovebox, and this solution was injected. Very slowly the vacuum line was opened to prevent bumping of the solution, and the solution was degassed until it reached a pressure of 100 mTorr. The solution was placed back under 5% hydrogen in argon. The temperature of the solution was then raised to 280 °C. In an argon glovebox, three separate vials of 1.540 g of *n*-hexadecane that had previously been heated and degassed for an hour at 90 °C at 60 mTorr was combined with a calculated amount of bis(trimethylsilyl)sulfide, a calculated amount of technical grade oleic acid that had previously been heated and degassed for an hour at 90 °C at 60 mTorr, and a calculated amount of diethyl zinc, respectively, and loaded into a syringe. The bis(trimethylsilyl)sulfide and diethyl zinc solutions were then injected into the flask using a syringe pump over the course of 1 hour, where 0.5 mL of the solution was injected in the first 45 minutes and the final 1.5 mL of solution was injected in the last 15 minutes, and simultaneously the oleic acid solution was injected in 0.1 mL increments every 3 minutes. The solution was allowed to stir for another 15 minutes. The flask was then rapidly cooled under a stream of nitrogen and hexanes. After cooling the solution was cannulated to be purified in an argon glovebox.

To determine the correct amount of each reagent, the radius and concentration of the indium phosphide quantum dots was calculated using the calculator provided by Ministro (74). Due to low yield of indium phosphide quantum dots, 3.5 mL of indium phosphide quantum dots was used (total yield was 4 mL) in the synthesis of the core/shell nanocrystals. Next, assuming that each monolayer of zinc sulfide was 0.269 nm thick and there are $7.97 \cdot 10^{18}$ zinc atoms/m$^2$ based on the (111) lattice constants and crystal structure of zinc sulfide (ICSD 52223), the number of sulfur atoms to cover the surface area of the indium phosphide quantum dots was computed. This added 0.134 nm to the radius, and then the number of zinc atoms to cover the surface area of the new nanocrystal was computed, adding another 0.134 nm. This process was repeated until the amount needed to add three layers of sulfur and three layers of zinc to the nanocrystals were computed. The total number of moles of zinc and sulfur was calculated and converted into a mass of diethyl zinc and bis(trimethylsilyl)sulfide needed. To calculate the amount of oleic acid need, the number of moles of diethyl zinc was multiplied by two and converted into a mass of oleic acid.

**Purification procedures for nanocrystal samples**

All washing steps were performed in an argon glovebox using 50 mL centrifuge tubes. For the metal sulfide nanocrystals and indium phosphide/zinc sulfide core/shell nanocrystal syntheses, to the synthesis solution methyl acetate was added in a 5:1 ratio. The solution was centrifuged for 5 minutes at 4562 RCF to form a pellet where the supernatant could be decanted. The pellet was resuspended in 4 mL of hexanes, and then 12 mL of methyl acetate was added to precipitate the nanocrystals again. After decanting the supernatant, the pellet was resuspended in 4 mL of tetrahydrofuran and then 12 mL of acetone was added to precipitate the nanocrystals. This step was necessary to remove free metal carboxylate ligands from the solution as it coordinates to metal centers and breaks up polymerized metal carboxylate structures (17, 75). The samples were then suspended in 4 mL of hexanes for future use.

For the indium phosphide quantum dots, the same steps as above were followed, but after suspension in tetrahydrofuran the solution would sit for one day before precipitation, and two more washing steps with tetrahydrofuran and acetone were needed, separated each by one day. This process ensured that no excess indium oleate was present in the final sample, which was suspended in hexanes.

For the cadmium oxide nanocrystals, after removing the solution to the glovebox, the solution was centrifuged at around 4562 RCF for five minutes. This caused all the nanocrystals to precipitate, after which 4 mL of hexanes was added to the pellet. Pellet was shaken and placed in a sonicating bath (sealed in a centrifuge tube) for five minutes. The solution was then centrifuged at 1711 RCF for two minutes, and the precipitate was discarded, leaving a black solution. This solution was washed two more times using 4 mL of hexanes and 8 mL of methyl acetate, using the sonicating bath when necessary to suspend the particles, before being suspended in hexanes and being centrifuged one last time at 1711 RC for two minutes. The supernatant was collected and saved.

**Synthesis of cadmium oleate**

The synthesis of cadmium oleate used in this study has previously been reported (76).

**Transmission electron microscopy and size measurements**

STEM imaging of the quantum dots was performed on the FEI TitanX 60-300 microscope at the National Center for Electron Microscopy, Lawrence Berkeley National Laboratory. STEM was performed at 300 kV with a beam convergence semiangle of 10 mrad using a Fischione HAADF detector with an inner collection half angle of 36 mrad. Bright Field TEM imaging of the nanocrystals was performed on an FEI Tecnai T20 S-TWIN TEM operated at 200 kV with a $LaB_6$ filament. Images were collected using a Gatan Rio 16IS camera with full 4k by 4k resolution. Ultrathin carbon TEM grids were used to increase contrast in the images. While some nanocrystal samples appeared fairly spherical, such as the zinc sulfide, cadmium sulfide, and lead sulfide quantum dots (*SI Appendix*, Fig. S2–4), which made for easy sizing and calculation of the radius, surface area, and volume of the nanoparticles, some indium phosphide samples, some indium phosphide/zinc sulfide core/shell samples, and the cadmium oxide samples were less spherical (*SI Appendix*, Fig. S1, S5–6). To accurately measure the surface area of all of these particles, regardless of how spherical they were, the projected areas were measured free hand. Using Cauchy's surface formula, it was assumed that the large number of nanoparticles measured were equivalent to random orientations of a convex, non-spherical object, and multiplying these areas by four gave an estimation of the surface area (77). From this, a radius and volume was then estimated. Although assumptions must be made, and not all of the nanocrystals were convex (some of the indium phosphide/zinc sulfide and cadmium oxide particles) based on the size distributions developed, the large number of particles analyzed, and the surface energy fitting, we conclude that this method is adequate. As the measurement of the enthalpy was done on an ensemble of particles, only the average molar surface area was important, and as nearly all size distributions appeared to follow a Gaussian distribution, the standard error in the average molar surface area is equal to the standard deviation of the molar surface area divided by the square root of the number of particles measured. However, for three of the indium phosphide/zinc sulfide core/shell nanocrystal samples and one of the cadmium oxide nanocrystal samples, the distribution was not Gaussian. In some of the core/shell samples it was evident that secondary nucleation of zinc sulfide had occurred, creating a second population of nanocrystals, and one sample of cadmium oxide had two populations. For the cadmium oxide samples, the actual molar surface area is relatively unimportant as the measured enthalpy does not significantly change with molar surface area, and so this was ignored. The analysis for the excess zinc sulfide nanocrystals is detailed below.

**Calorimetry experiments**

Calorimetry measurements were performed on a Setaram C80 calorimeter with custom built quartz reaction cells. The sample and reference cells were filled with 5 mL of 37% hydrochloric acid each and were maintained at a temperature of 25 °C. After approximately 10 hours of equilibration, samples were dropped into the sample chamber and the heat flow recorded. Measurements were typically completed and the heat flow returned to a linear baseline after approximately 10 hours, depending on the sample. For the measurement of bulk materials, samples were dropped as pieces or chunks, while for the nanocrystal samples, solutions of the colloidal nanocrystals were dropcast onto a Teflon platform from hexanes until approximately 5–10 mg had been deposited, and then the sample was placed under a vacuum at 60 mTorr for 30 minutes before the final mass deposited was recorded. Background measurements of the Teflon platform dropped into 37% hydrochloric acid were performed, measuring at an average of 600 mJ with a standard deviation of 165 mJ, and these measurements were used for the error on individual measurements from the calorimeter. Raw thermograms of the measurements can be found in *SI Appendix*, Fig. S8-14. An interesting observation from the measurements in Fig. 1B and 1C is that the heat flow for both measurements is initially exothermic, but the magnitude of the signal is much greater for the quantum dot sample. This is consistent with the dissolution of indium phosphide occurring at the surface and, considering the much higher surface area of the quantum dots, a larger heat flow is to be expected. However, the dissolution rapidly becomes endothermic for the quantum dot samples. Unfortunately, it is difficult to directly assign the cause of this switch to an endothermic signal as it could be due to a different reaction that may be occurring at a different rate, or it could be compensation by the instrument for overcorrection of the heat flow to maintain an isothermal temperature. The oleate ligands are protonated to create oleic acid, but as oleic acid is not soluble in aqueous solutions, it does not dissolve in the hydrochloric acid. If the signals are integrated over time, the total heat of the reaction, or the enthalpy of the reaction, can be calculated.

**CHNS elemental analysis**

Elemental analysis was performed on isolated samples of dried nanocrystals and cadmium oleate using a Perkin Elmer 2400 Series II combustion analyzer. Results for the dried nanocrystals are found in *SI Appendix*, Table S1, while the cadmium oleate yielded 63.98% carbon and 9.62% hydrogen, and the theoretical carbon and hydrogen content is 64.00% and 9.90%, respectively. The experimental uncertainty on these measurements is 0.3%. Determination of oleate content for the nanocrystal samples was done by dividing the carbon content by 0.7681, the fraction of oleate that is carbon. For metal sulfide nanocrystals, these measurements were sufficient to determine the elemental content of the samples.

**Inductively coupled plasma optical emission spectroscopy measurements**

Indium phosphide and indium phosphide/zinc sulfide samples were dried of solvent and dissolved in Optima for Ultra Trace Elemental Analysis 67–69% nitric acid. Standard solutions of indium, phosphorus, zinc, and cadmium were made at concentrations of 25, 5, 1, 0.5, and 0.1 ppm. The solutions were analyzed on a Perkin Elmer ICP Optima 7000 DV Spectrometer, and the ratio of indium, phosphorus, and zinc could be computed and are found in *SI Appendix*, Table S1 with their determined experimental uncertainties. Coupled with the CHNS elemental analysis, these measurements were sufficient to determine the fraction of all elements in the indium phosphide and indium phosphide/zinc sulfide samples. The cadmium oxide nanocrystals were analyzed quantitatively by drying and weighing the sample before dissolution in a known quantity of nitric acid, and the concentration of cadmium ions could be measured directly. With the CHNS elemental analysis, these measurements were sufficient to determine the elemental content of the samples.

**Analysis of zinc sulfide quantum dot contamination**

As mentioned above, there was some contamination of zinc sulfide quantum dots in three of the indium phosphide/zinc sulfide core/shell nanocrystal samples. However, as the enthalpy of zinc sulfide quantum dots synthesized in a similar fashion was computed in this study, it was relatively simple to correct for this heat effect. Using the molar masses estimated for the size series of zinc sulfide and indium phosphide, the mass fraction of the zinc sulfide quantum dots was calculated and based on the size and surface area the heat of the dissolution of these quantum dots was subtracted from the total measurement, which resulted in a small number (see *SI Appendix*, Table S2).

**Nuclear magnetic resonance characterization**

Solution measurements on nanocrystal samples were performed on a Bruker Avance 700 instrument. Samples were dried and then suspended in toluene-$d_8$. Organic ligands that were free in the solution have sharper peaks than those bound to nanocrystals and are upshifted (17), see *SI Appendix*, Fig. S1–6 for examples of the vinyl protons on ligands bound to nanocrystals and *SI Appendix*, Fig. S7 for examples of oleic acid and cadmium oleate that are free in solution. Measurements on oleic acid samples were done on a Bruker Avance NEO 500 while the measurement on the synthesized cadmium oleate was performed on a Bruker Avance 400 instrument. Note the broadness of the cadmium oleate vinyl peak and its slightly shifted position, this is likely due to it being in the form of polymerized metal carboxylate (75). Also note the acidic proton peak in *SI Appendix*, Fig. S7A compared with Fig. S7B. The isolated oleic acid after the reaction of the quantum dots with hydrochloric acid has a broader and more shifted acidic proton peak around 12 ppm due to water picked up from the reaction, but it does show the formation of oleic acid.

Solid state magic-angle spinning NMR was performed by packing dried nanocrystal samples into $ZrO_2$ rotors and spinning at 40,000 RPM under a flow of nitrogen gas on a Bruker Avance 500 instrument. The lack of peaks around 0 ppm shows that there is no detectable oxidation in these samples (*SI Appendix*, Fig. S1 and S6). Additionally, the lower signal from the indium phosphide/zinc sulfide core/shell nanocrystals reflects the lower phosphorus content in these samples, and the lack of hardly any detectable phosphorus signal in the cross-polarized measurement (*SI Appendix*, Fig. S6F) shows that the distance between phosphorus atoms and hydrogen atoms is large, indicating the growth of a zinc sulfide shell that coats the indium phosphide core.

**Optical measurements**

The absorbance spectra of dilute samples of nanocrystals in hexanes in a 1 cm pathlength quartz cuvette were taken on a Shizmadu UV-3600 spectrometer using a slit width of 1.0 nm and a scan speed of 350 nm/min at a resolution of 1.0 nm and were baseline corrected using a cuvette filled with hexanes. Note that beyond approximately 1560 nm the instrument signal was lost, hence why one absorption spectra of the largest lead sulfide quantum dots does not return to baseline (*SI Appendix*, Fig. S4). Luminescence measurements on the indium phosphide and indium phosphide/zinc sulfide core/shell samples were taken on a Horiba Jobin Yvon TRIAX 320 Fluorolog. Excitation wavelength was 437 nm with an excitation and emission slit width of 2.5 nm taken with a resolution of 1.0 nm, integrated for 20 seconds. Luminescence measurements on the zinc sulfide, cadmium sulfide, and lead sulfide quantum dots was taken on an Edinburgh FLS 980 Spectrometer with variable excitation wavelengths depending on the sample and an excitation and emission slit width of 2 nm taken with a resolution of 1.0 nm, integrated for 2 seconds. Absolute quantum yield measurements for the indium phosphide and indium phosphide/zinc sulfide core/shell nanocrystals were performed by a custom built spectrometer, the details of which have been published elsewhere (78).

**Powder X-ray diffraction measurements**

Quantum dot solutions were drop-cast onto a silicon wafer from hexanes and the X-ray diffraction patterns were taken on a Bruker D2 Phaser with a copper K-alpha source (wavelength 0.1541 nm). The spectra were collected from 10° to 70° 2θ with a constant φ angle rotation of 72°/s. The broad peak around 20° 2θ in all measurements, except the cadmium oxide nanocrystals, in *SI Appendix*, Tables S1–4 and S6 has been previously identified as ordered ligands on the surface of colloidal quantum dots, most obvious on the smallest quantum dots, hence why it is not visible on the cadmium oxide nanocrystals (10). The bulk measurement on cadmium sulfide shows a mixture of zinc blende and wurtzite phases (*SI Appendix*, Fig. S3B) but considering that the enthalpy of formation of both phases are nearly identical (79), this phase impurity is fortunately a non-issue.

**Acknowledgments**

This work was supported by the U.S. Department of Energy, Office of Science, Office of Basic Energy Sciences, Materials Sciences and Engineering Division, under Contract No. DE-AC02-05-CH11231 (Physical Chemistry of Inorganic Nanostructures Program (KC3103)). J.J.C. and M.F.C. gratefully acknowledge the National Science Foundation Graduate Research Fellowship under Grant DGE 1752814. J.J.C. also acknowledges support by the Kavli NanoScience Institute, University of California, Berkeley through the Philomathia Graduate Student Fellowship. Work at the Molecular Foundry was supported by the Office of Science, Office of Basic Energy Sciences, of the U.S. Department of Energy under Contract No. DE-AC02-05CH11231. We would like to than Eran Rabani, Dipti Jasrasaria, and Kaiyue Peng for helpful conversations regarding surface and interface energy calculations, and we would also like to thank Alexandra Navrotsky for guidance on instrumentation and schematics for the reaction vessels as well as discussion providing justification for neglecting surface entropy. We thank College of Chemistry's NMR facility for resources provided and Alicia Lund for her assistance. Instruments in CoC-NMR are supported in part by NIH S10OD024998. We would also like to thank Elena Kreimer for performing the CHNS elemental analysis.

**Figures and Tables**

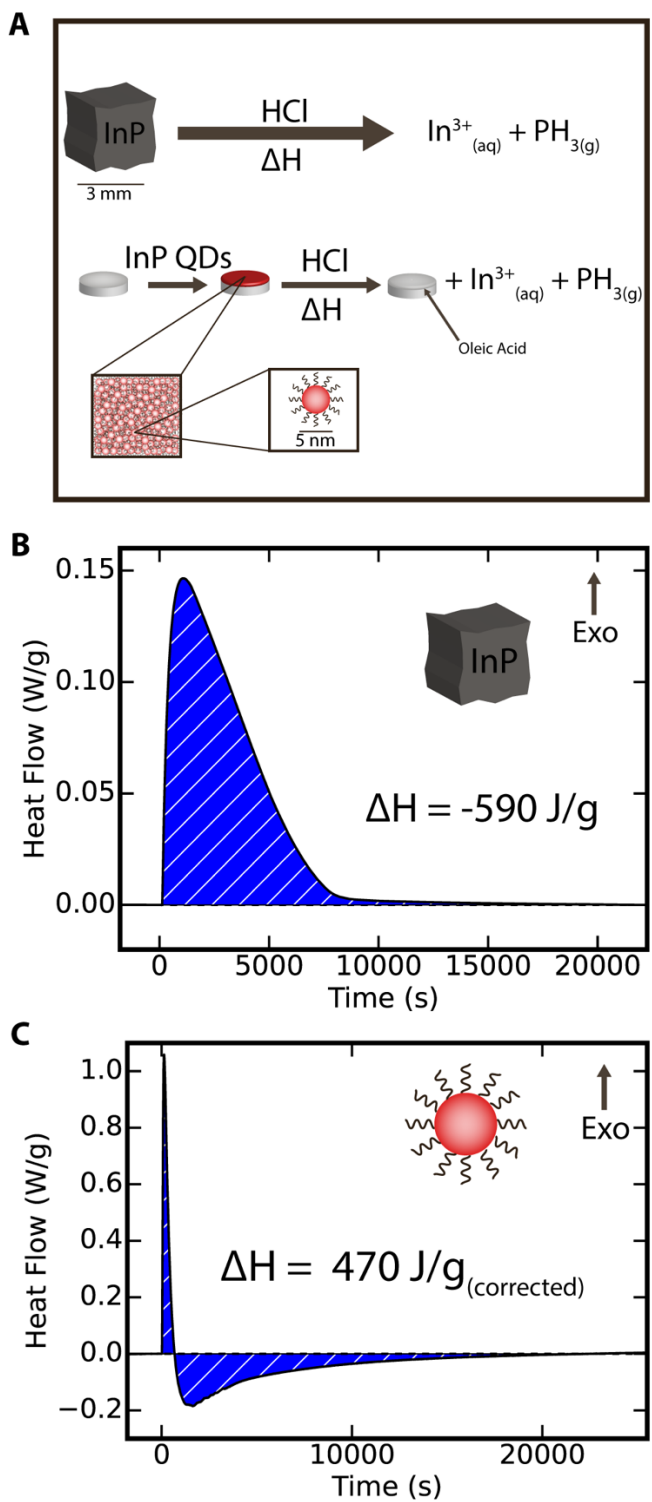

**Figure 1.** (A) Reaction equations for the dissolution of bulk (top) and quantum dots (bottom) of indium phosphide in hydrochloric acid. The indium phosphide quantum dots are deposited on a Teflon platform as an agglomeration. After dissolution of the indium phosphide quantum dots,

oleic acid is left behind. (B) Heat flow measurements of the reaction of bulk indium phosphide with hydrochloric acid over time, with the integrated area in blue and the total enthalpy recorded as -590 J/g. Note that a positive heat flow is an exothermic reaction, while a negative heat flow is an endothermic reaction. (C) Heat flow measurements of the reaction of quantum dots of indium phosphide with hydrochloric acid over time, with the integrated area in blue and the total corrected enthalpy recorded as 470 J/g. Note that the thermal effects of the Teflon platform and protonation of oleate must be corrected for, hence the corrected enthalpy.

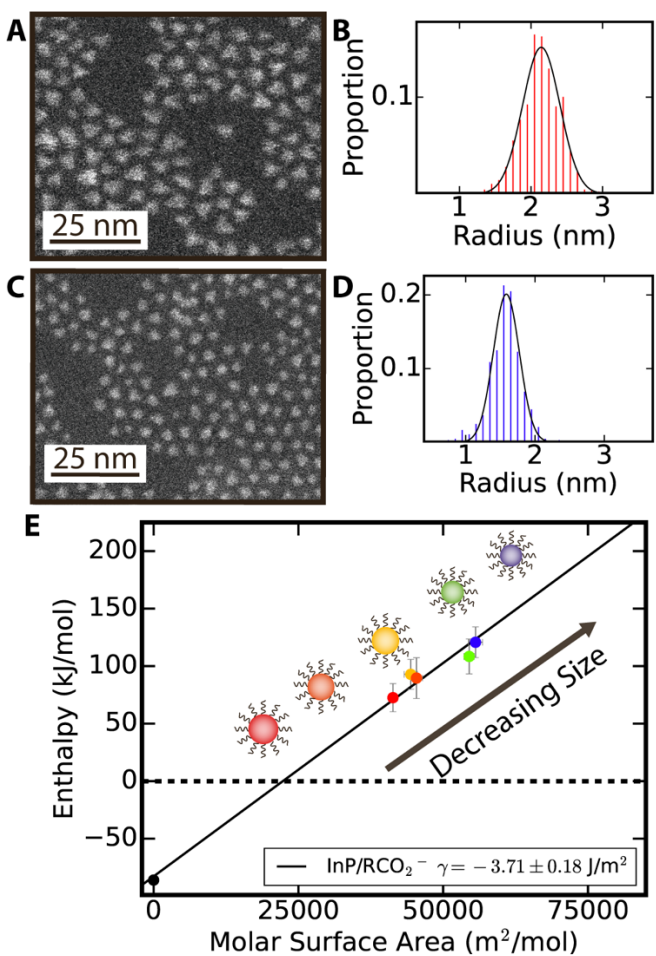

**Figure 2.** (A) Scanning transmission electron microscope (STEM) image of the largest indium phosphide quantum dot sample used in these measurements with inset scale bar. (B) Histogram fit with a Gaussian distribution of the largest indium phosphide quantum dot sample used in these measurements. (C) STEM image of the smallest indium phosphide quantum dot sample used in these measurements with inset scale bar. (D) Histogram fit with a Gaussian distribution of the smallest indium phosphide quantum dot sample used in these measurements. (E) Plot of the enthalpy measured in kJ/mol against the molar surface area in $m^2$/mol for the indium phosphide samples with a fit line where the negative slope of the fit line is the measured surface energy of the indium phosphide quantum dots (-3.71 ± 0.18 $J/m^2$) with cartoon representations of the indium phosphide quantum dots included. As molar surface area increases, the size of the quantum dots decreases. Error in enthalpy measured is propagated standard error from the calorimetry measurements as well as measurements of the sample mass and molar mass and calculated protonation enthalpy (see *SI Appendix*, Table S2). Error in molar surface area measured is propagated standard error from the sizing measurements (see *SI Appendix*, Table S1 and S2 and Materials and Methods section for detailed calculations and methodology). The error reported in the surface energy is the standard deviation calculated from a least-squares fit of the datapoints.

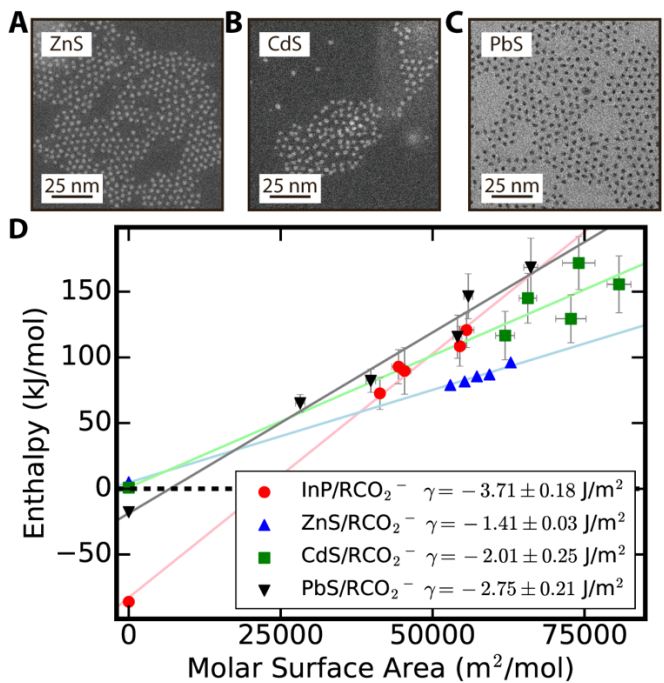

**Figure 3.** (A) Scanning transmission electron microscope (STEM) image of a sample of zinc sulfide quantum dots used in these measurements with inset scale bar. (B) STEM image of a sample of cadmium sulfide quantum dots used in these measurements with inset scale bar. (C) Transmission electron microscope image of a sample of lead sulfide quantum dots used in these measurements with inset scale bar. (D) Plot of the enthalpy measured in kJ/mol against the molar surface area in $m^2$/mol for the different material samples with fits line for the different samples where the negative slope of the fit line is the measured surface energy of the quantum dots, and the measured surface energy for each material is reported in the legend. As molar surface area increases, the size of the quantum dots decreases. Error in enthalpy measured is propagated standard error from the calorimetry measurements as well as measurements of the sample mass and molar mass and calculated protonation enthalpy (see *SI Appendix*, Table S2). Error in molar surface area measured is propagated standard error from the sizing measurements (see *SI Appendix*, Table S1 and S2 and Materials and Methods section for detailed calculations and methodology). The error reported in the surface energies is the standard deviation calculated from a least-squares fit of the datapoints.

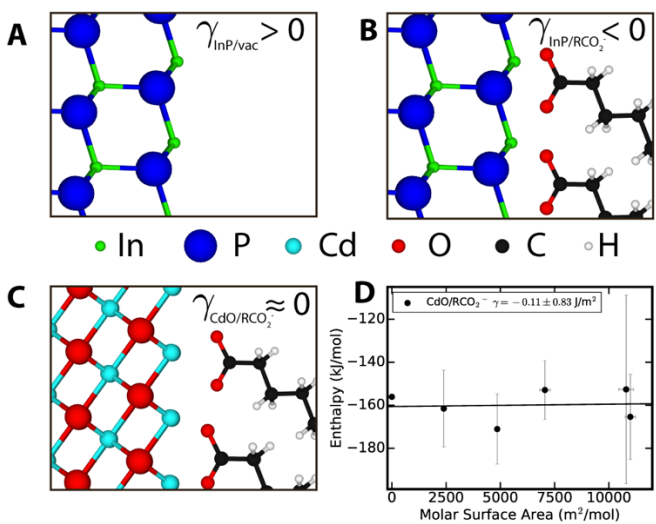

**Figure 4.** (A) Model of the surface of indium phosphide in vacuum, where the terminating indiums have fewer bonds than the interior indiums and thus have a positive surface energy. (B) Model of the surface of indium phosphide terminated with coordinating carboxylates, and if these bonds are stronger than the interior bonds, the surface energy will be negative. (C) Model of the surface of cadmium oxide terminated with coordinating carboxylates, where the interior and exterior oxygen bonds have similar enthalpies, resulting in a near zero surface energy. (D) Plot of the enthalpy measured in kJ/mol against the molar surface area in $m^2$/mol for the cadmium oxide samples with a fit line where the negative slope of the fit line is the measured surface energy of the cadmium oxide nanocrystals (-0.11 ± 0.83 $J/m^2$). As molar surface area increases, the size of the nanocrystals decreases. Error in enthalpy measured is propagated standard error from the calorimetry measurements as well as measurements of the sample mass and molar mass and calculated protonation enthalpy (see *SI Appendix*, Table S2). Error in molar surface area measured is propagated standard error from the sizing measurements (see *SI Appendix*, Table S1 and S2 and Materials and Methods section for detailed calculations and methodology). The error reported in the surface energies is the standard deviation calculated from a least-squares fit of the datapoints.

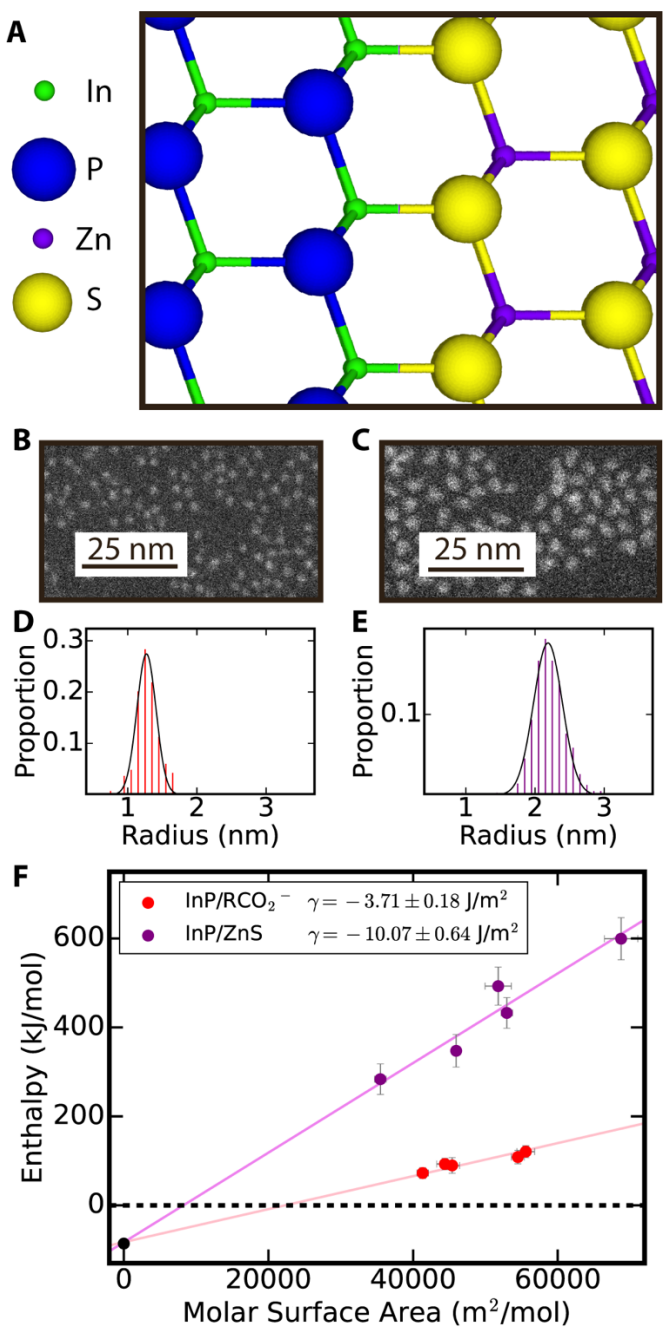

**Figure 5.** (A) Model of the interface of indium phosphide and zinc sulfide core/shell nanocrystals, where sulfur atoms coordinate to indium atoms at the interface. (B) Scanning transmission electron microscope (STEM) image of the smallest indium phosphide quantum dot sample that was shelled with zinc sulfide with inset scale bar. (C) Histogram fit with a Gaussian distribution of the smallest indium phosphide quantum dot sample that was shelled with zinc sulfide. (D) STEM image of the smallest indium phosphide quantum dot sample after shelling with zinc sulfide used in these measurements with inset scale bar. (E) Histogram fit with a Gaussian distribution of the smallest indium phosphide quantum dot sample after shelling with zinc sulfide used in these measurements. (F) Plot of the enthalpy measured in kJ/mol against the molar surface area in

m$^2$/mol for the indium phosphide/zinc sulfide core/shell samples with a fit line where the negative slope of the fit line is the measured interfacial energy of the indium phosphide/zinc sulfide core/shell samples (-10.07 ± 0.64 J/m$^2$) alongside the indium phosphide/carboxylate surface energy for comparison. As molar surface area increases, the size of the core nanocrystals decreases. Error in enthalpy measured is propagated standard error from the calorimetry measurements as well as measurements of the sample mass and molar mass and calculated protonation enthalpy, zinc sulfide enthalpy, and zinc sulfide surface energy (see *SI Appendix*, Table S2). Error in molar surface area measured is propagated standard error from the sizing measurements (see SI Appendix, Table S1 and S2 and Materials and Methods section for detailed calculations and methodology). The error reported in the surface energy is the standard deviation calculated from a least-squares fit of the datapoints.

**Supporting Information for**
Observation of negative surface and interface energies of quantum dots

Jason J. Calvin, Amanda S. Brewer, Michelle F. Crook, Tierni M. Kaufman, and A. Paul Alivisatos

A. Paul Alivisatos
Email: paul.alivisatos@berkeley.edu

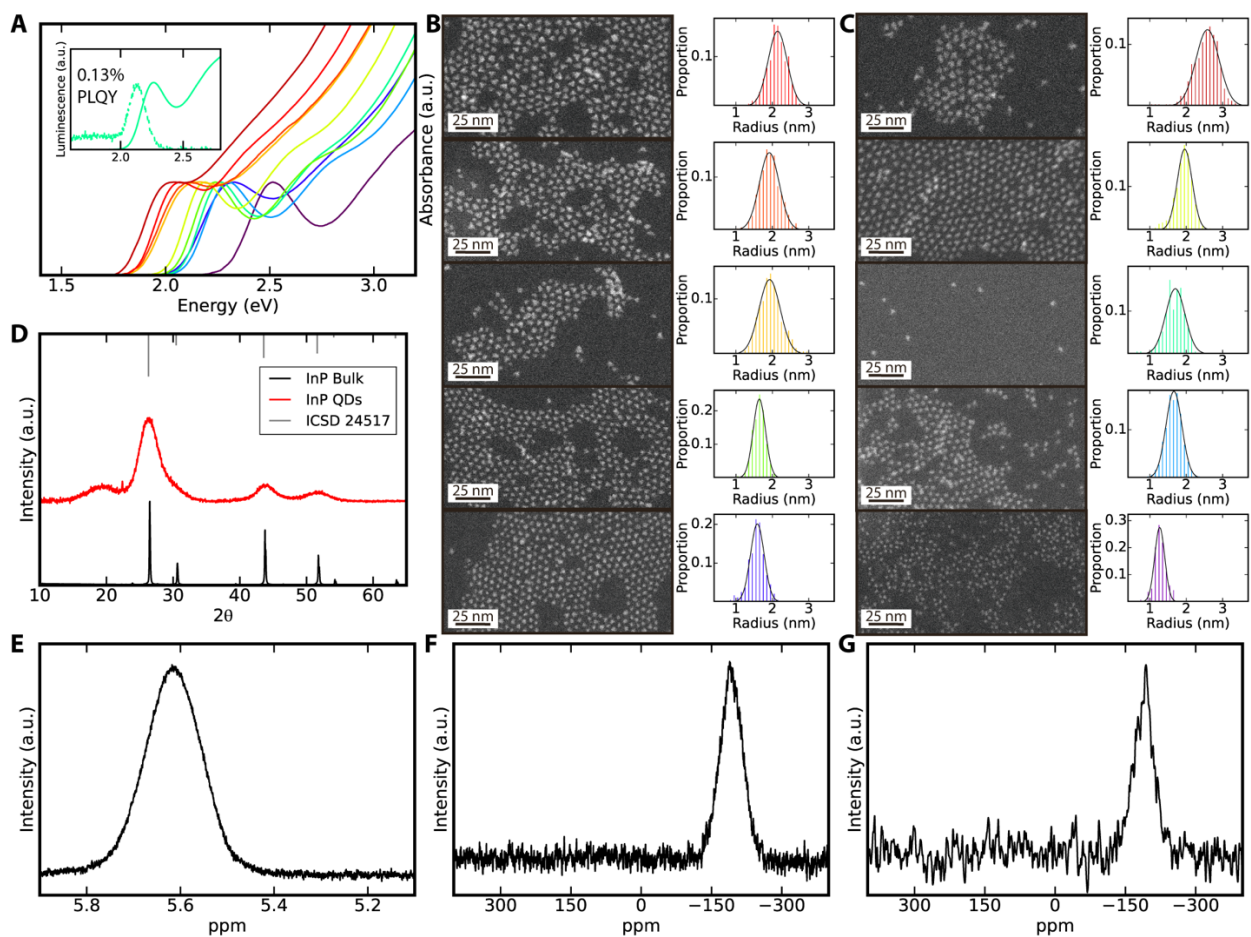

**Fig. S1.** (A) Absorbance measurements of the ten indium phosphide quantum dot samples used herein with inset including absorbance and luminescence measurements for one representative sample with the measured photoluminescence quantum yield (PLQY). (B) Scanning transmission electron microscope (STEM) images of the indium phosphide quantum dot samples used for the indium phosphide/carboxylate surface energy measurements with inset scale bar to the left, and matched histograms fit with by Gaussian distributions to the left. (C) STEM images of the indium phosphide quantum dot samples to be shelled with zinc sulfide with inset scale bars to the left, and matched histograms fit by Gaussian distributions to the right. (D) Powder X-ray diffraction measurements of bulk and a representative quantum dot sample of indium phosphide with the reference lines for zinc blende indium phosphide from the inorganic crystal structure database (ICSD 24517). (E) $^1$H NMR measurements of the vinyl proton region of a representative sample of indium phosphide quantum dots in toluene-$d_8$. (F) Solid state $^{31}$P magic-angle spinning measurements of a representative sample of indium phosphide quantum dots. (G) Solid state cross-polarized $^1$H-$^{31}$P magic-angle spinning measurements of a representative sample of indium phosphide quantum dots.

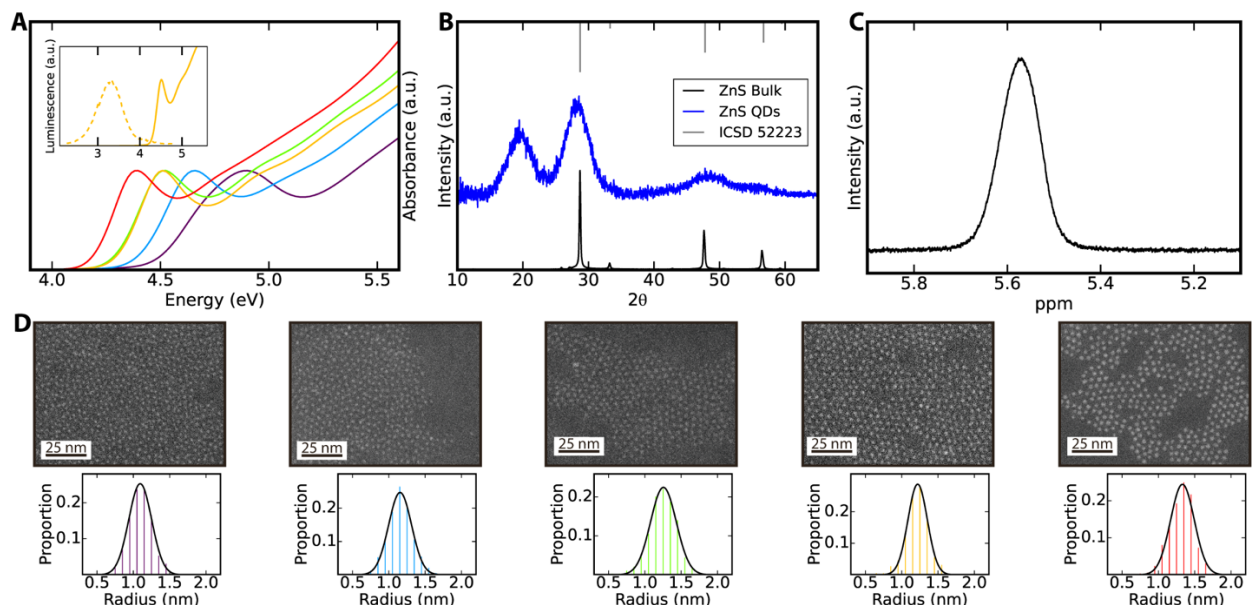

**Fig. S2.** (A) Absorbance measurements of the five zinc sulfide quantum dot samples used herein with inset including absorbance and luminescence measurements for one representative sample. (B) Powder X-ray diffraction measurements of bulk and a representative quantum dot sample of zinc sulfide with the reference lines of zinc blende zinc sulfide from the inorganic crystal structure database (ICSD 52223). (C) $^1$H NMR measurements of the vinyl proton region of a representative sample of zinc sulfide quantum dots in toluene-$d_8$. (D) Scanning transmission electron microscope images of the zinc sulfide quantum dot samples used for the zinc sulfide/carboxylate surface energy measurements with inset scale bars to the top, and matched histograms fit by Gaussian distributions to the bottom.

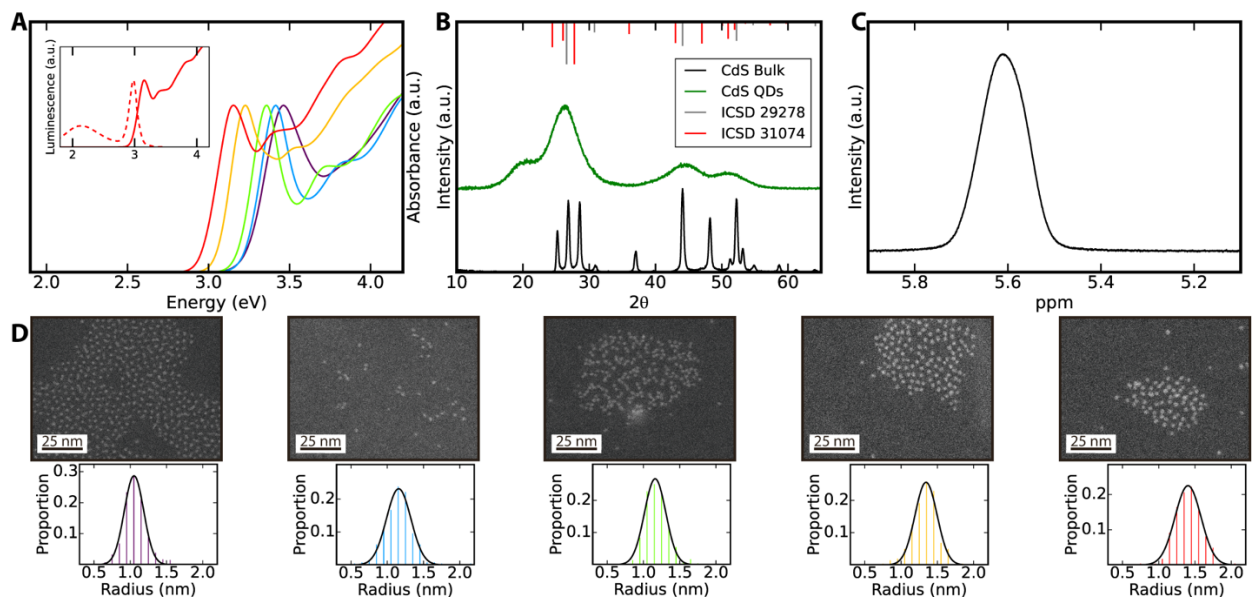

**Fig. S3.** (A) Absorbance measurements of the five cadmium sulfide quantum dot samples used herein with inset including absorbance and luminescence measurements for one representative sample. (B) Powder X-ray diffraction measurements of bulk and a representative quantum dot sample of zinc sulfide with the reference lines of both zinc blende and wurtzite cadmium sulfide from the inorganic crystal structure database (ICSD 29278 and 31074, respectively). (C) $^1$H NMR measurements of the vinyl proton region of a representative sample of cadmium sulfide quantum dots in toluene-$d_8$. (D) Scanning transmission electron microscope images of the cadmium sulfide quantum dot samples used for the cadmium sulfide/carboxylate surface energy measurements with inset scale bars to the top, and matched histograms fit by Gaussian distributions to the bottom.

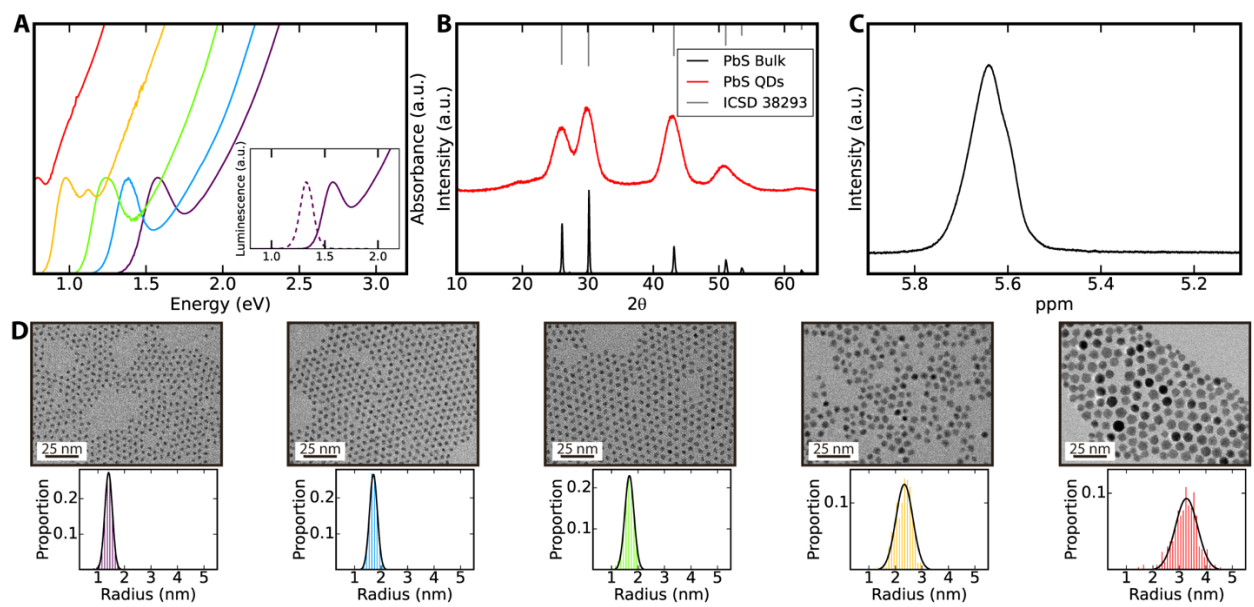

**Fig. S4.** (A) Absorbance measurements of the five lead sulfide quantum dot samples used herein with inset including absorbance and luminescence measurements for one representative sample. (B) Powder X-ray diffraction measurements of bulk and a representative quantum dot sample of lead sulfide with the reference lines of rock salt lead sulfide from the inorganic crystal structure database (ICSD 38293). (C) $^1$H NMR measurements of the vinyl proton region of a representative sample of lead sulfide quantum dots in toluene-$d_8$. (D) Transmission electron microscope images of the lead sulfide quantum dot samples used for the lead sulfide/carboxylate surface energy measurements with inset scale bars to the top, and matched histograms fit by Gaussian distributions to the bottom.

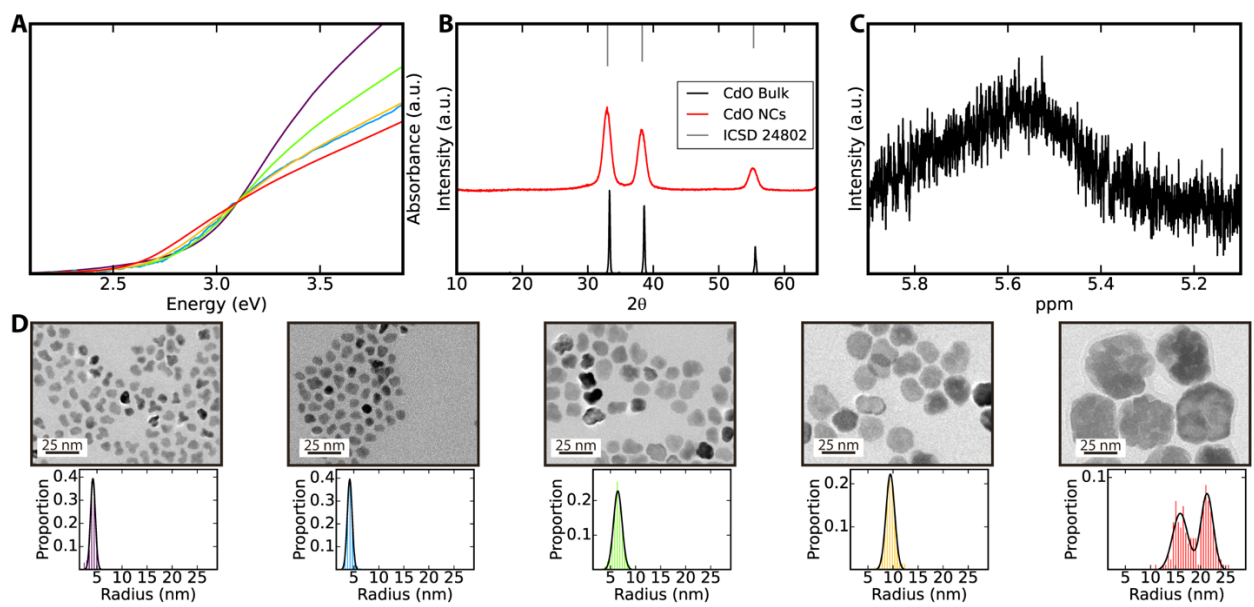

**Fig. S5.** (A) Absorbance measurements of the five cadmium oxide nanocrystal samples used herein. (B) Powder X-ray diffraction measurements of bulk and a representative nanocrystal sample of cadmium oxide with the reference lines of rock salt cadmium oxide from the inorganic crystal structure database (ICSD 24802). (C) $^1$H NMR measurements of the vinyl proton region of a representative sample of cadmium oxide quantum dots in toluene-$d_8$. (D) Transmission electron microscope images of the cadmium oxide nanocrystal samples used for the cadmium oxide/carboxylate surface energy measurements with inset scale bars to the top, and matched histograms fit by Gaussian distributions to the bottom.

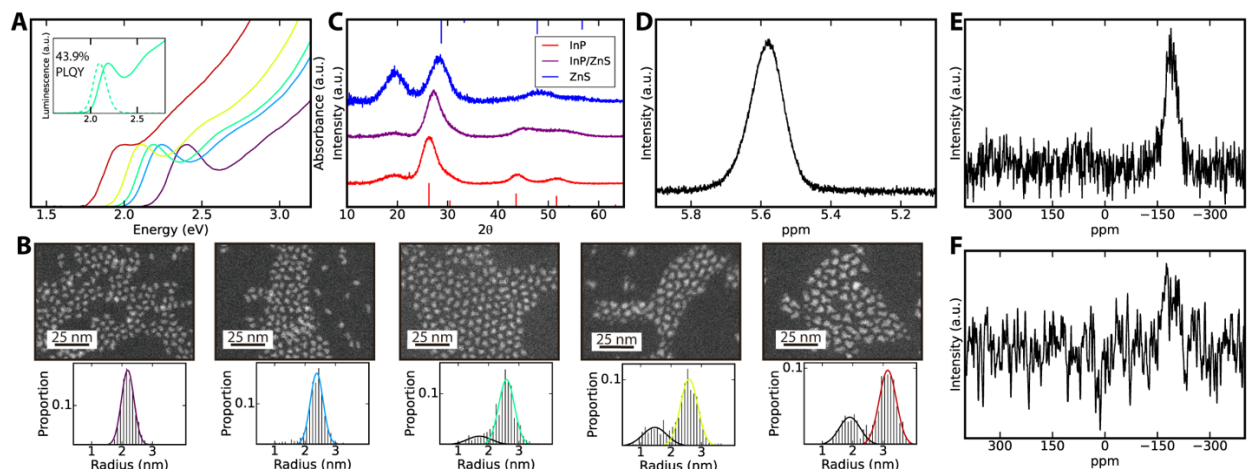

**Fig. S6.** (A) Absorbance measurements of the five indium phosphide/zinc sulfide core/shell nanocrystal samples used herein with inset including absorbance and luminescence measurements for one representative sample with the measured photoluminescence quantum yield (PLQY). (B) Scanning transmission electron microscope (STEM) images of the indium phosphide/zinc sulfide core/shell nanocrystal samples used for the indium phosphide/zinc sulfide interface energy measurements with inset scale bar to the top, and matched histograms fit with by Gaussian distributions to the bottom. (C) Powder X-ray diffraction measurements of representative nanocrystal samples of indium phosphide, zinc sulfide, and indium phosphide/zinc sulfide with the reference lines for zinc blende indium phosphide and zinc blende zinc sulfide from the inorganic crystal structure database (ICSD 24517 and 52223, respectively). (D) $^1$H NMR measurements of the vinyl proton region of a representative sample of indium phosphide/zinc sulfide core/shell nanocrystals in toluene-$d_8$. (E) Solid state $^{31}$P magic-angle spinning measurements of a representative sample of indium phosphide/zinc sulfide core/shell nanocrystals. (F) Solid state cross-polarized $^1$H-$^{31}$P magic-angle spinning measurements of a representative sample of indium phosphide/zinc sulfide core/shell nanocrystals.

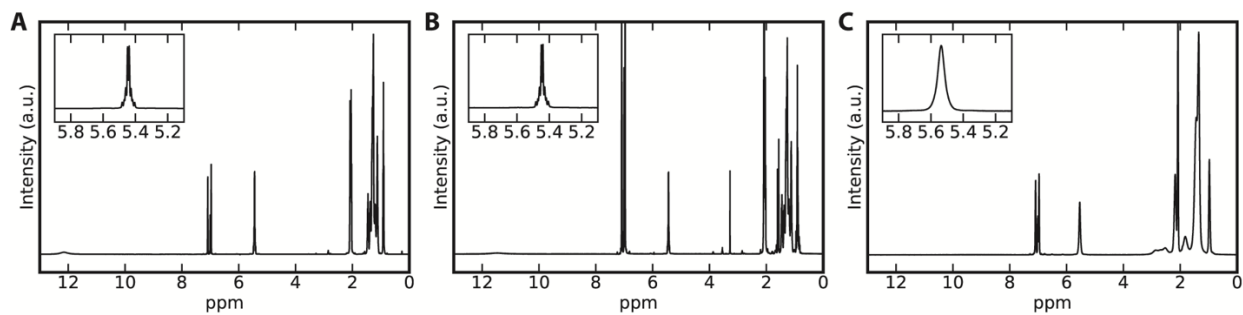

**Fig. S7.** (A) $^1$H NMR spectra of 99% oleic acid in toluene-$d_8$ with inset of the vinyl proton region. (B) $^1$H NMR spectra of recovered oleic acid from the calorimetry experiments in toluene-$d_8$ with inset of the vinyl proton region. (C) $^1$H NMR spectra of synthesized cadmium oleate in toluene-$d_8$ with inset of the vinyl proton region. Note solvent peaks at 7.09, 7.00, 6.98, and 2.09 ppm as well as a peak attributed to water in the spectra of the recovered oleic acid at 3.28 ppm.

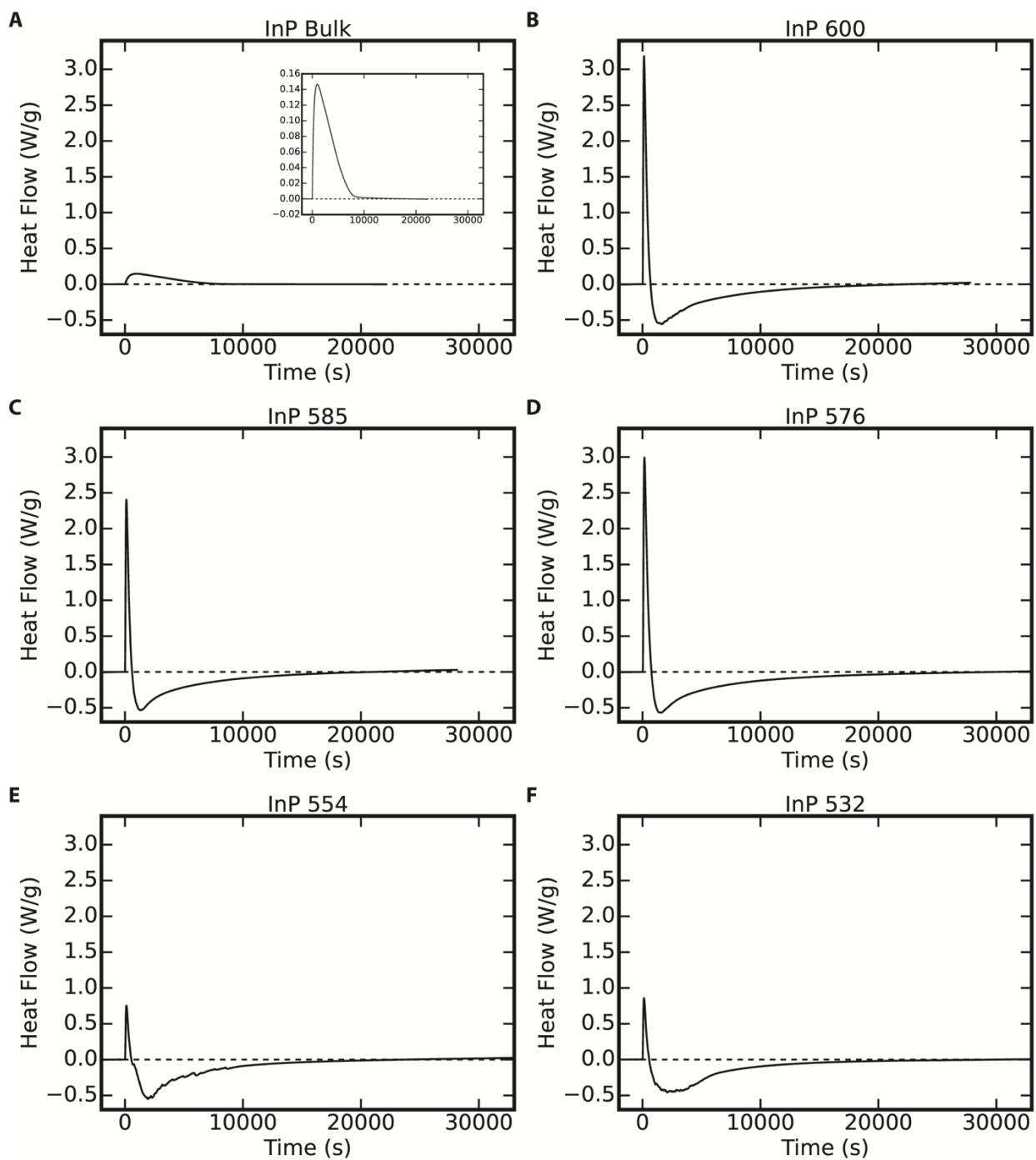

**Fig. S8.** (A) Raw dissolution thermogram of bulk indium phosphide. (B) Raw dissolution thermogram of InP 600. (C) Raw dissolution thermogram of InP 585. (D) Raw dissolution thermogram of InP 576. (E) Raw dissolution thermogram of InP 554. (F) Raw dissolution thermogram of InP 532. Samples are given labels based on the wavelength in nanometers of the first exciton peak in the absorption spectrum.

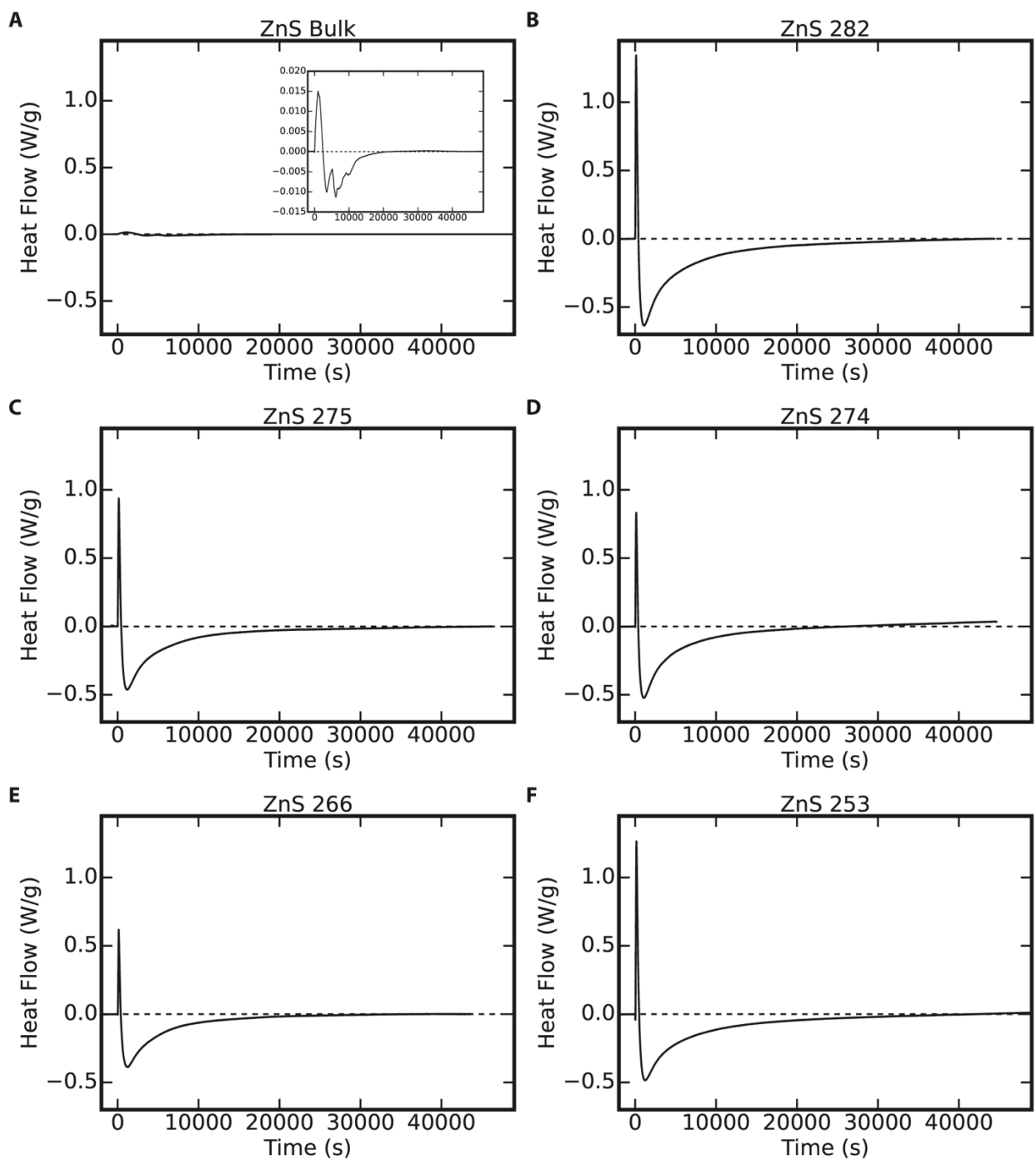

**Fig. S9.** (A) Raw dissolution thermogram of bulk zinc sulfide. (B) Raw dissolution thermogram of ZnS 282. (C) Raw dissolution thermogram of ZnS 275. (D) Raw dissolution thermogram of ZnS 274. (E) Raw dissolution thermogram of ZnS 266. (F) Raw dissolution thermogram of ZnS 253. Samples are given labels based on the wavelength in nanometers of the first exciton peak in the absorption spectrum.

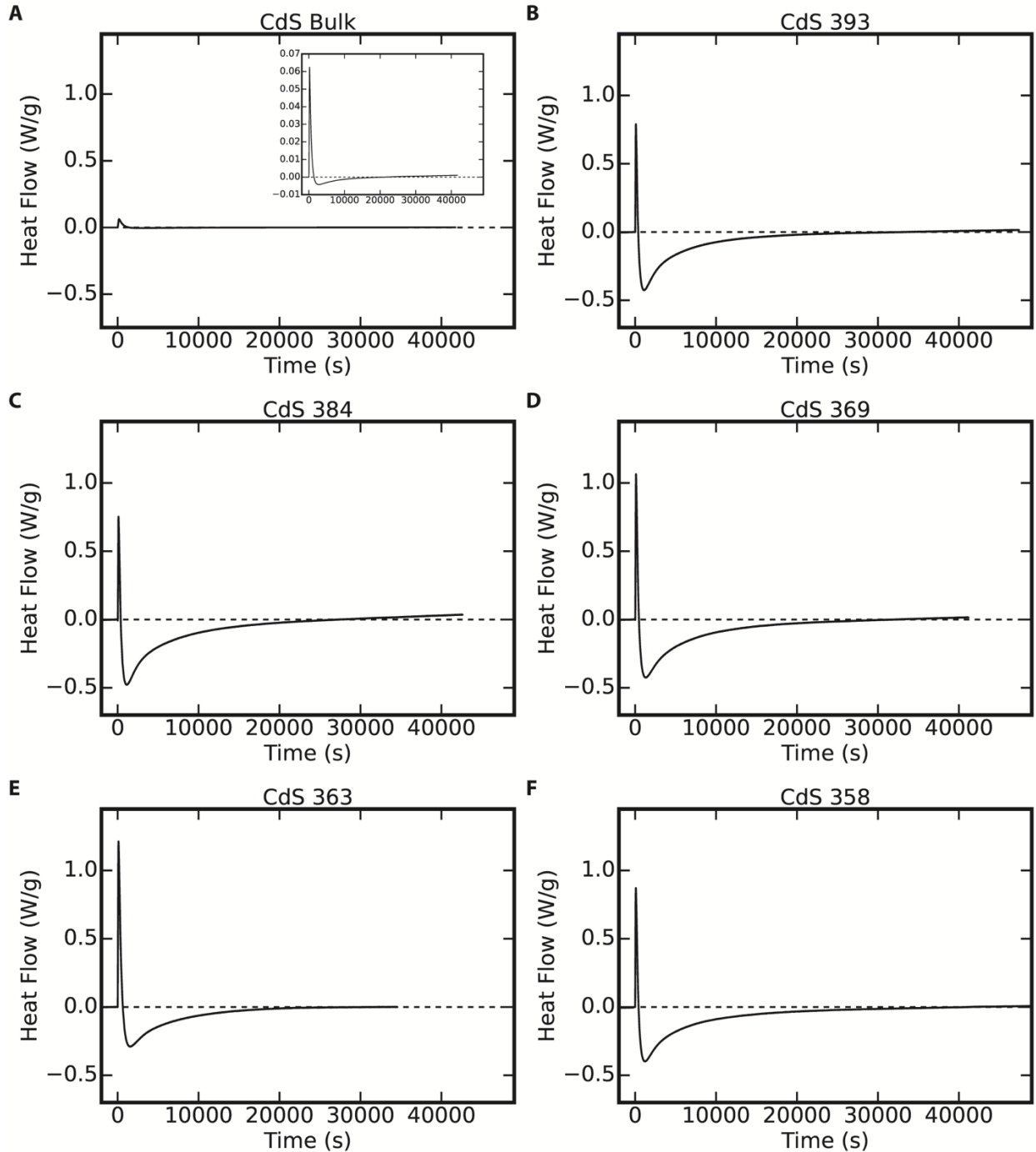

**Fig. S10.** (A) Raw dissolution thermogram of bulk cadmium sulfide. (B) Raw dissolution thermogram of CdS 393. (C) Raw dissolution thermogram of CdS 384. (D) Raw dissolution thermogram of CdS 369. (E) Raw dissolution thermogram of CdS 363. (F) Raw dissolution thermogram of CdS 358. Samples are given labels based on the wavelength in nanometers of the first exciton peak in the absorption spectrum.

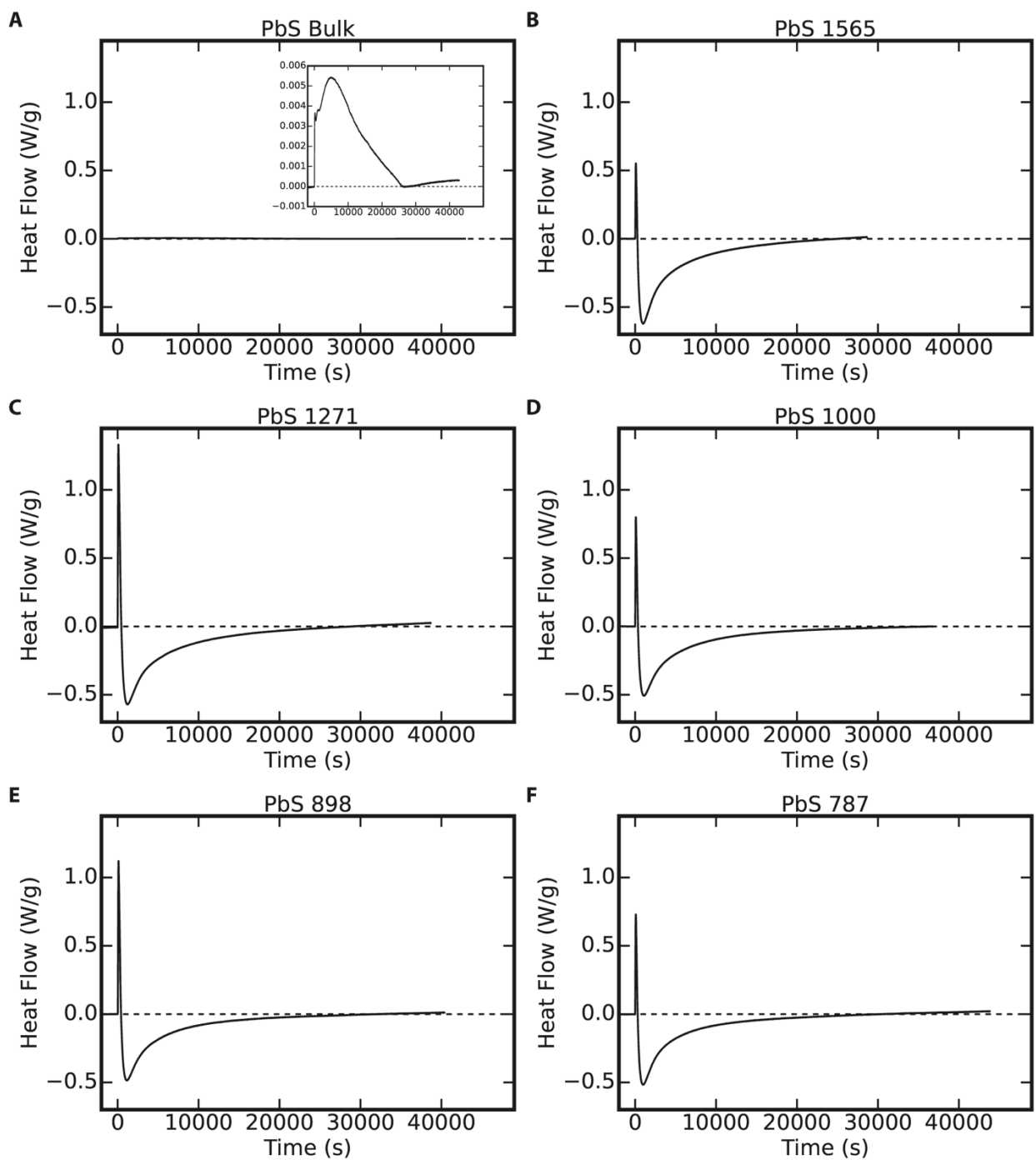

**Fig. S11.** (A) Raw dissolution thermogram of bulk lead sulfide. (B) Raw dissolution thermogram of PbS 1565. (C) Raw dissolution thermogram of PbS 1271. (D) Raw dissolution thermogram of PbS 1000. (E) Raw dissolution thermogram of PbS 898. (F) Raw dissolution thermogram of PbS 787. Samples are given labels based on the wavelength in nanometers of the first exciton peak in the absorption spectrum.

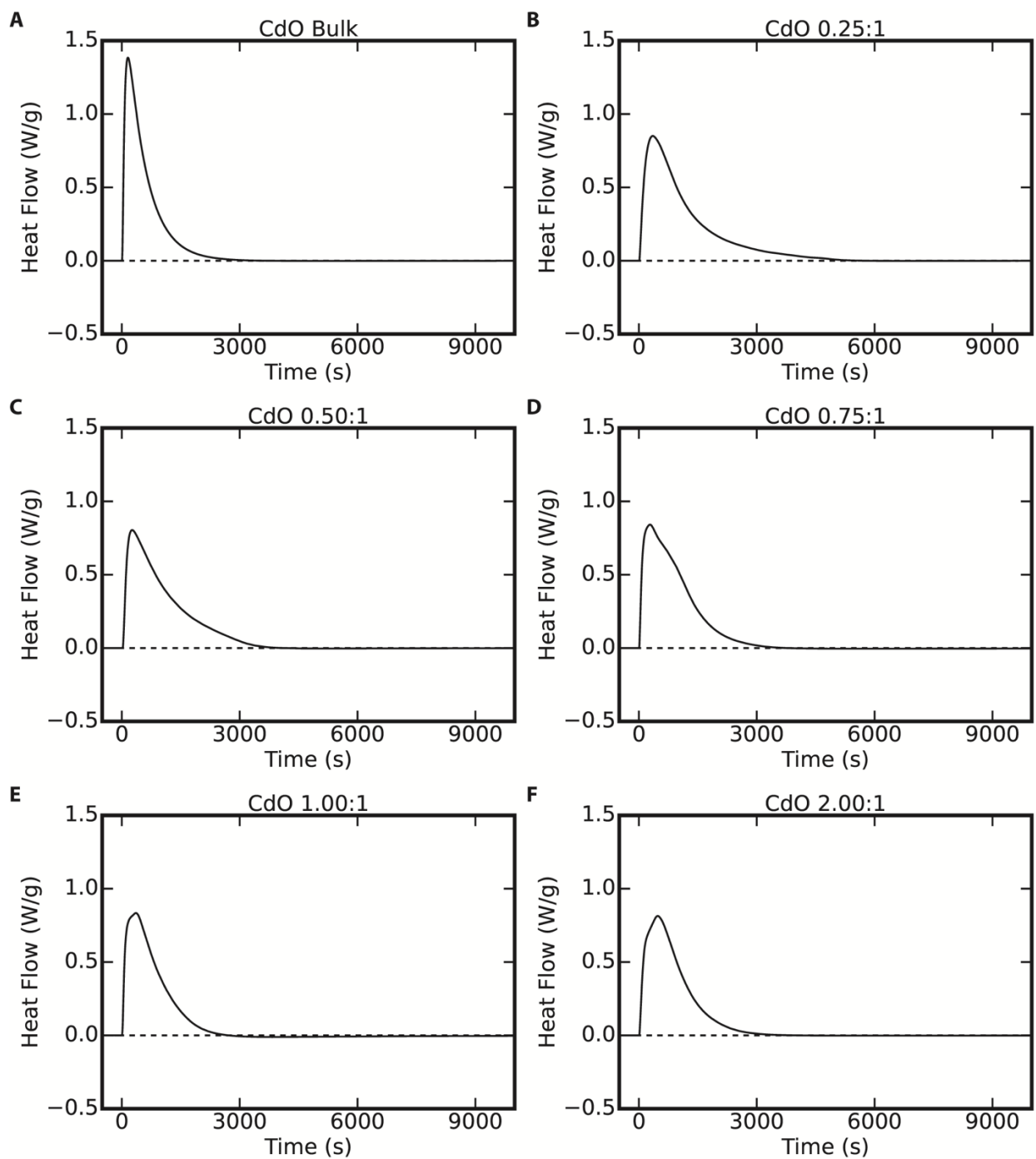

**Fig. S12.** (A) Raw dissolution thermogram of bulk cadmium oxide. (B) Raw dissolution thermogram of CdO 0.25:1. (C) Raw dissolution thermogram of CdO 0.50:1. (D) Raw dissolution thermogram of CdO 0.75:1. (E) Raw dissolution thermogram of CdO 1.00:1. (F) Raw dissolution thermogram of CdO 2.00:1. Samples are given labels based on the ratio of bis(trimethylsilyl)amine to cadmium acetylacetonate.

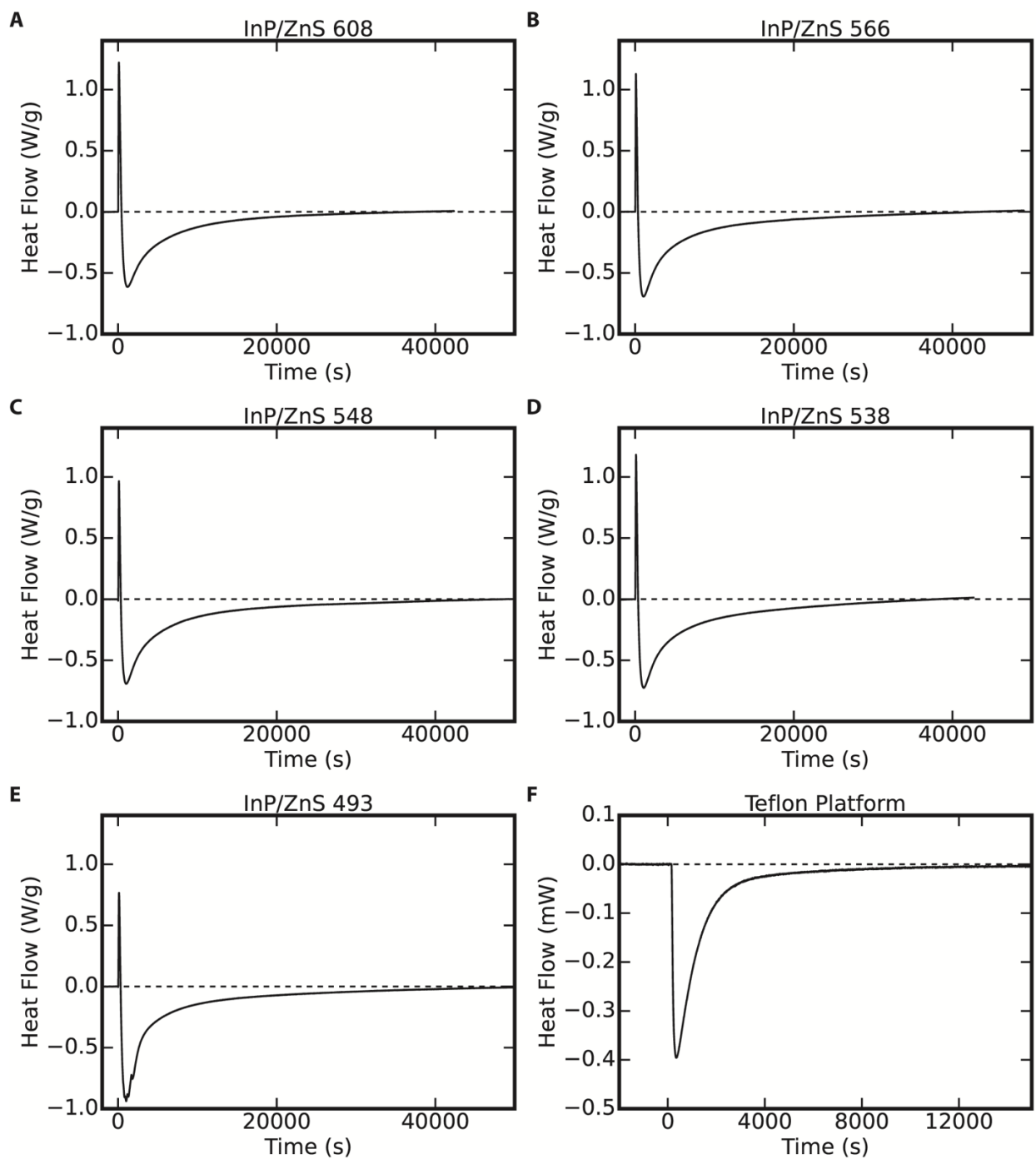

**Fig. S13.** (A) Raw dissolution thermogram of InP/ZnS 608. (B) Raw dissolution thermogram of InP/ZnS 566. (C) Raw dissolution thermogram of InP/ZnS 548. (D) Raw dissolution thermogram of InP/ZnS 538. (E) Raw dissolution thermogram of InP/ZnS 493. (F) Raw dissolution thermogram of the clean Teflon platform. Samples are given labels based on the wavelength in nanometers of the first exciton peak in the absorption spectrum of the indium phosphide cores.

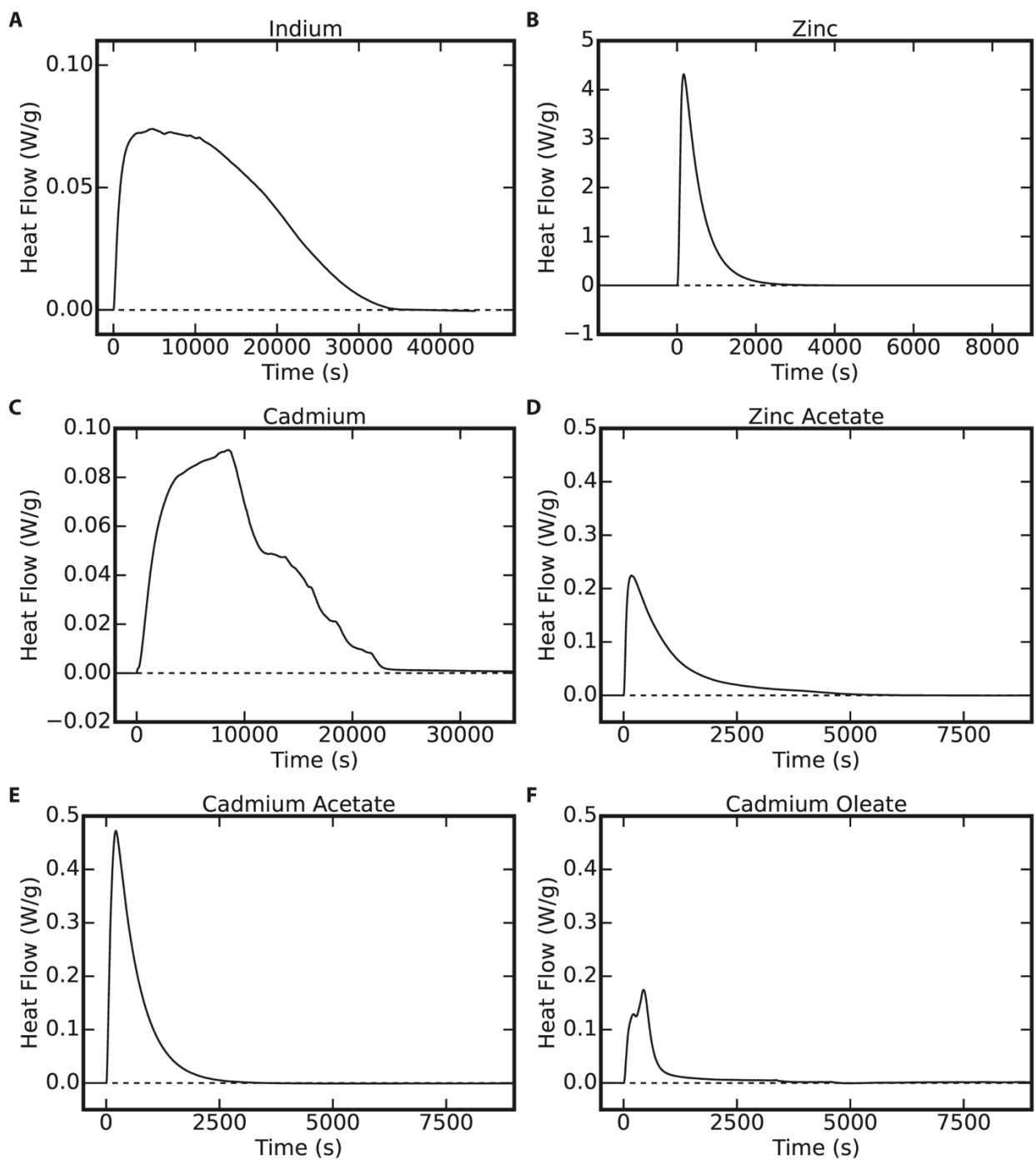

**Fig. S14.** (A) Raw dissolution thermogram of indium. (B) Raw dissolution thermogram of zinc. (C) Raw dissolution thermogram of cadmium. (D) Raw dissolution thermogram of zinc acetate. (E) Raw dissolution thermogram of cadmium acetate. (F) Raw dissolution thermogram of cadmium oleate.

**Table S1.** Characterization of samples used in calorimetry experiments. Samples are given labels based on the wavelength in nanometers of the first exciton peak in the absorption spectrum except in the case of cadmium oxide nanocrystals, and instead the ratio of bis(trimethylsilyl)amine to cadmium acetylacetonate is used. Count refers to the number of nanocrystals whose area was measured. The results of the CHNS elemental analysis for each element is reported, with an assumed error of 0.3%. Note that the indium phosphide samples for shelling with zinc sulfide were not measured by CHNS elemental analysis. Ligand coverage was calculated from the radius of particles and CHNS elemental analysis, and due to zinc sulfide contamination in some indium phosphide zinc sulfide samples, ligand coverages were not calculated for these samples.

| Sample | Count | Radius (nm) | Surface Area (nm$^2$) | Volume (nm$^3$) | C% | H% | N% | S% | Ligand Coverage (nm$^{-2}$) | In:P | In:Zn | Zn:S | Cd:S | Pb:S | Cd% |
|---|---|---|---|---|---|---|---|---|---|---|---|---|---|---|---|
| InP 532 | 500 | 1.57±0.01 | 31.8±0.4 | 17.3±0.3 | 39.75 | 5.83 | - | - | 6.5±0.3 | 1.34±0.01 | - | - | - | - | - |
| InP 554 | 501 | 1.63±0.008 | 33.9±0.3 | 18.9±0.3 | 34.92 | 5.38 | - | - | 5.2±0.8 | 1.40±0.09 | - | - | - | - | - |
| InP 576 | 502 | 1.95±0.01 | 49.1±0.7 | 33.5±0.7 | 38.27 | 5.82 | - | - | 7.5±0.7 | 1.28±0.04 | - | - | - | - | - |
| InP 585 | 501 | 1.93±0.01 | 47.6±0.6 | 31.8±0.6 | 33.93 | 5.03 | - | - | 5.8±1.3 | 1.3±0.1 | - | - | - | - | - |
| InP 600 | 512 | 2.14±0.01 | 58.2±0.6 | 42.7±0.7 | 35.26 | 5.36 | - | - | 6.8±0.6 | 1.22±0.03 | - | - | - | - | - |
| InP 493 | 170 | 1.28±0.01 | 21.0±0.4 | 9.3±0.3 | - | - | - | - | - | - | - | - | - | - | - |
| InP 538 | 1087 | 1.656±0.007 | 35.1±0.3 | 20.1±0.2 | - | - | - | - | - | - | - | - | - | - | - |
| InP 548 | 215 | 1.68±0.02 | 36.4±0.7 | 21.3±0.6 | - | - | - | - | - | - | - | - | - | - | - |
| InP 566 | 1000 | 1.919±0.008 | 47.1±0.4 | 31.0±0.3 | - | - | - | - | - | - | - | - | - | - | - |
| InP 608 | 506 | 2.55±0.02 | 83±1 | 73±1 | - | - | - | - | - | - | - | - | - | - | - |
| InP/ZnS 493 | 518 | 2.21±0.01 | 61.7±0.5 | 46.3±0.6 | 29.04 | 4.42 | - | 11.20 | - | 1.35±0.05 | 1.35±0.05 | 1.86±0.06 | - | - | - |
| InP/ZnS 538 | 503 | 2.34±0.01 | 70.2±0.7 | 56.5±0.9 | 29.36 | 4.46 | - | 8.85 | - | 1.28±0.03 | 1.28±0.03 | 2.14±0.08 | - | - | - |
| InP/ZnS 548 | 504 | 2.59±0.01 | 85.2±0.8 | 75±1 | 33.52 | 4.98 | - | 8.11 | - | 1.24±0.05 | 1.24±0.05 | 2.17±0.09 | - | - | - |
| InP/ZnS 566 | 905 | 2.59±0.01 | 85.6±0.7 | 76±1 | 27.22 | 4.24 | - | 10.59 | - | 1.26±0.07 | 1.26±0.07 | 1.89±0.06 | - | - | - |
| InP/ZnS 608 | 504 | 3.15±0.02 | 125±1 | 134±2 | 24.87 | 3.71 | - | 13.04 | - | 1.11±0.03 | 1.11±0.03 | 1.66±0.05 | - | - | - |
| ZnS 253 | 500 | 1.094±0.007 | 15.3±0.2 | 5.8±0.1 | 52.46 | 8.15 | - | 6.77 | 7.9±0.7 | - | - | 1.81±0.09 | - | - | - |
| ZnS 266 | 505 | 1.156±0.008 | 17.1±0.2 | 6.9±0.1 | 49.04 | 7.56 | - | 8.32 | 6.7±0.5 | - | - | 1.64±0.07 | - | - | - |
| ZnS 274 | 510 | 1.244±0.008 | 19.8±0.2 | 8.6±0.2 | 48.96 | 7.57 | - | 9.09 | 7.1±0.5 | - | - | 1.47±0.06 | - | - | - |
| ZnS 275 | 500 | 1.211±0.007 | 18.7±0.2 | 7.8±0.1 | 47.06 | 7.30 | - | 9.23 | 6.2±0.5 | - | - | 1.57±0.06 | - | - | - |
| ZnS 282 | 834 | 1.312±0.006 | 22.0±0.2 | 9.9±0.1 | 43.02 | 6.64 | - | 11.02 | 5.3±0.4 | - | - | 1.47±0.05 | - | - | - |
| CdS 358 | 509 | 1.069±0.007 | 14.7±0.2 | 5.5±0.1 | 47.64 | 7.30 | - | 5.12 | 7.3±0.8 | - | - | - | 1.8±0.1 | - | - |
| CdS 363 | 295 | 1.15±0.01 | 17.2±0.3 | 7.0±0.2 | 47.63 | 7.38 | - | 5.48 | 7.8±0.9 | - | - | - | 1.7±0.1 | - | - |
| CdS 369 | 317 | 1.185±0.009 | 18.0±0.3 | 7.4±0.2 | 46.36 | 7.07 | - | 5.67 | 7.4±0.9 | - | - | - | 1.71±0.09 | - | - |
| CdS 384 | 373 | 1.330±0.009 | 22.6±0.3 | 10.3±0.2 | 44.26 | 6.76 | - | 6.35 | 7.2±0.7 | - | - | - | 1.62±0.08 | - | - |
| CdS 393 | 312 | 1.41±0.01 | 25.4±0.4 | 12.3±0.3 | 42.34 | 6.54 | - | 6.98 | 6.9±0.7 | - | - | - | 1.55±0.07 | - | - |
| PbS 787 | 500 | 1.394±0.007 | 24.7±0.2 | 11.8±0.2 | 32.35 | 4.96 | - | 5.11 | 6.6±1.1 | - | - | - | - | 1.6±0.01 | - |
| PbS 896 | 584 | 1.717±0.007 | 37.4±0.3 | 21.8±0.2 | 29.77 | 4.56 | - | 5.84 | 6.8±1.1 | - | - | - | - | 1.47±0.08 | - |
| PbS 1000 | 814 | 1.654±0.006 | 34.8±0.3 | 19.6±0.2 | 28.22 | 4.33 | - | 6.29 | 6.0±1.0 | - | - | - | - | 1.40±0.07 | - |
| PbS 1271 | 501 | 2.30±0.01 | 67.4±0.8 | 53.3±0.9 | 22.78 | 3.60 | - | 7.66 | 5.9±1.1 | - | - | - | - | 1.27±0.05 | - |
| PbS 1565 | 298 | 3.23±0.03 | 133±2 | 149±3 | 15.30 | 2.44 | - | 9.42 | 4.8±1.2 | - | - | - | - | 1.16±0.04 | - |
| CdO 2.00:1 | 457 | 4.14±0.03 | 220±3 | 316±6 | 14.28 | 2.08 | - | - | 6.0±4.9 | - | - | - | - | - | 72.6±0.9 |
| CdO 1.00:1 | 208 | 4.24±0.04 | 230±4 | 335±9 | 11.60 | 1.83 | - | - | 5.0±12.3 | - | - | - | - | - | 77±2 |
| CdO 0.75:1 | 221 | 6.43±0.06 | 530±10 | 1180±30 | 6.61 | 0.97 | - | - | 3.8±5.3 | - | - | - | - | - | 80.6±0.8 |
| CdO 0.50:1 | 394 | 9.53±0.05 | 1150±10 | 3740±60 | 8.36 | 1.23 | - | - | 7.6±8.3 | - | - | - | - | - | 80.3±0.7 |
| CdO 0.25:1 | 175 | 18.8±0.2 | 4500±100 | 30000±1000 | 4.41 | 0.65 | - | - | 7.2±19.6 | - | - | - | - | - | 83±1 |

**Table S2.** Calorimetry data for the nanocrystal samples. The mass recorded has an error of 0.01 mg, and mass of oleate (OA) was determined by CHNS elemental analysis. For the three largest indium phosphide/zinc sulfide core/shell nanocrystal samples that had detectable zinc sulfide contamination, the calculated mass of the excess zinc sulfide material is reported. The calculated error for the enthalpy measurements is 165 mJ.

| Sample | Mass (mg) | Mass OA (mg) | Mass ZnS QDs (mg) | Enthalpy (mJ) | Protonation Enthalpy (mJ) | ZnS Shell (mJ) | ZnS QDs (mJ) | Enthalpy (J/g) | MW (g/mol) | Molar Surface Area (m²/mol) | Enthalpy (kJ/mol) |
|---|---|---|---|---|---|---|---|---|---|---|---|
| InP 532 | 5.80 | 3.00±0.02 | - | 2800 | 78±7 | - | - | 760±80 | 158±2 | 56000±1000 | 120±10 |
| InP 554 | 5.12 | 2.33±0.02 | - | 2600 | 60±5 | - | - | 680±80 | 160±10 | 54500±900 | 110±20 |
| InP 576 | 5.78 | 2.88±0.02 | - | 2400 | 75±6 | - | - | 600±80 | 156±5 | 44000±1000 | 90±10 |
| InP 585 | 4.05 | 1.79±0.02 | - | 1900 | 46±4 | - | - | 600±100 | 160±10 | 45000±1000 | 90±20 |
| InP 600 | 5.56 | 2.55±0.02 | - | 2100 | 66±6 | - | - | 470±80 | 154±4 | 41300±800 | 70±10 |
| InP/ZnS 493 | 4.99 | 1.89±0.02 | - | 5400 | 49±4 | 1020±50 | - | 3800±300 | 158±6 | 69000±2000 | 600±50 |
| InP/ZnS 538 | 4.97 | 1.90±0.02 | - | 4800 | 49±4 | 990±40 | - | 2800±200 | 156±4 | 53000±800 | 430±30 |
| InP/ZnS 548 | 4.84 | 2.11±0.02 | 0.111±0.009 | 4700 | 55±5 | 820±40 | 70±6 | 3200±200 | 155±6 | 52000±2000 | 490±40 |
| InP/ZnS 566 | 5.09 | 1.80±0.02 | 0.23±0.01 | 4400 | 47±4 | 950±50 | 159±8 | 2200±200 | 155±8 | 46000±600 | 350±40 |
| InP/ZnS 608 | 4.61 | 1.49±0.02 | 0.27±0.01 | 3600 | 39±3 | 760±30 | 147±7 | 1900±200 | 150±4 | 35500±800 | 280±30 |
| ZnS 253 | 8.43 | 5.79±0.03 | - | 3200 | 150±10 | - | - | 900±90 | 107±5 | 63000±1000 | 100±10 |
| ZnS 266 | 4.67 | 2.98±0.02 | - | 2100 | 77±7 | - | - | 800±100 | 106±4 | 59000±1000 | 90±20 |
| ZnS 274 | 6.06 | 3.86±0.02 | - | 2400 | 100±9 | - | - | 800±100 | 104±4 | 55000±1000 | 80±10 |
| ZnS 275 | 6.01 | 3.68±0.02 | - | 2600 | 95±8 | - | - | 800±100 | 105±4 | 57000±1000 | 90±10 |
| ZnS 282 | 9.19 | 5.15±0.04 | - | 3800 | 130±10 | - | - | 760±60 | 104±3 | 53900±800 | 79±7 |
| CdS 358 | 5.36 | 3.32±0.02 | - | 2600 | 86±7 | - | - | 900±100 | 170±10 | 81000±2000 | 160±20 |
| CdS 363 | 5.85 | 3.63±0.02 | - | 3000 | 94±8 | - | - | 1000±100 | 165±9 | 74000±3000 | 170±20 |
| CdS 369 | 5.86 | 3.54±0.02 | - | 2500 | 92±8 | - | - | 800±100 | 166±9 | 73000±2000 | 130±20 |
| CdS 384 | 5.18 | 2.98±0.02 | - | 2600 | 77±7 | - | - | 900±100 | 163±8 | 66000±1000 | 150±20 |
| CdS 393 | 4.79 | 2.64±0.02 | - | 2200 | 68±6 | - | - | 700±100 | 162±7 | 62000±2000 | 120±20 |
| PbS 787 | 5.72 | 2.41±0.02 | - | 2700 | 62±5 | - | - | 600±70 | 280±20 | 66000±1000 | 170±20 |
| PbS 896 | 6.83 | 2.64±0.03 | - | 2400 | 69±6 | - | - | 420±60 | 270±10 | 54100±700 | 120±20 |
| PbS 1000 | 6.40 | 2.35±0.03 | - | 2900 | 61±5 | - | - | 550±60 | 270±10 | 55900±700 | 150±20 |
| PbS 1271 | 10.55 | 3.13±0.04 | - | 3000 | 81±7 | - | - | 320±30 | 260±10 | 39800±600 | 82±9 |
| PbS 1565 | 11.86 | 2.36±0.05 | - | 3100 | 61±5 | - | - | 260±10 | 252±8 | 28200±800 | 65±7 |
| CdO 2.00:1 | 8.02 | 1.49±0.03 | - | -7300 | 39±3 | - | - | -1220±40 | 140±20 | 11000±200 | -170±20 |
| CdO 1.00:1 | 5.00 | 0.76±0.02 | - | -4000 | 20±2 | - | - | -1080±60 | 140±40 | 10800±300 | -150±40 |
| CdO 0.75:1 | 7.59 | 0.65±0.03 | - | -7500 | 17±2 | - | - | -1160±30 | 130±10 | 7100±200 | -150±10 |
| CdO 0.50:1 | 6.02 | 0.66±0.02 | - | -5900 | 17±2 | - | - | -1210±40 | 140±10 | 4860±90 | -170±20 |
| CdO 0.25:1 | 12.95 | 0.74±0.05 | - | -14400 | 19±2 | - | - | -1230±20 | 130±10 | 2400±100 | -160±20 |

**Table S3.** Calorimetry data for bulk materials measured and reference data used. The calculated error for the enthalpy measurements is 165 mJ.

| Material | Mass (mg) | Enthalpy (mJ) | Enthalpy (kJ/mol) | Experimental $\Delta H_f$ (kJ/mol) | Reference $\Delta H_f$ (kJ/mol) | Reference |
|---|---|---|---|---|---|---|
| $In_{(s)}$ | 167.7±0.1 | -248600 | -170.2±0.2 | - | - | - |
| $Zn_{(s)}$ | 42.1±0.1 | -106300 | -165.1±0.5 | - | - | - |
| $Cd_{(s)}$ | 178.1±0.1 | -200700 | -126.6±0.1 | - | - | - |
| $PH_{3(g)}$ | - | - | - | - | 5.47±0.4 | (1) |
| $H_2S_{(g)}$ | - | - | - | - | -20.6±0.5 | (2) |
| $H_2O_{(l)}$ | - | - | - | - | -285.83±0.04 | (2) |
| $InP_{(s)}$ | 243.6±0.1 | -143700 | -86.0±0.1 | -78.7±0.4 | -69±3 | (3) |
| $ZnS_{(s)}$ | 40.6±0.1 | 2100 | 5.0±0.4 | -190.7±0.8 | -207±4 | (4) |
| $CdS_{(s)}$ | 75.4±0.1 | 400 | 0.8±0.3 | -148.0±0.6 | -148±4 | (4) |
| $PbS_{(s)}$ | 37.2±0.1 | -2800 | -18±1 | - | -98±4 | (4) |
| $CdO_{(s)}$ | 62.3±0.1 | -75600 | -155.8±0.4 | -256.6±0.5 | -255.7±2 | (5) |
| $Zn(OAc)_{2(s)}$ | 71.29±0.01 | -17100 | -43.9±0.4 | - | -1078.6±0.1 | (6) |
| $Cd(OAc)_{2(s)}$ | 20.22±0.01 | -6400 | -73±2 | -1011±2 | - | - |
| $Cd(OA)_{2(s)}$ | 6.47±0.01 | -700 | -70±20 | - | - | - |
| $OAc^-_{(aq)}$ | - | - | - | - | -486.01±0.1 | (7) |
| $HOAc_{(aq)}$ | - | - | - | -478.7±0.3 | - | - |

**SI References**